\documentclass[
  journal=pasa,
  manuscript=review, 
  year=2022,
  volume=37,
]{cup-journal}

\usepackage{microtype,booktabs}
\usepackage[utf8]{inputenc}
\usepackage{bm}
\usepackage{amsmath}
\usepackage{mathtools}
\usepackage[english]{babel}
\usepackage{breqn}
\usepackage{siunitx}
\DeclareSIUnit\gauss{G}
\usepackage[title]{appendix}
\usepackage{tikz}
\usepackage{tikz-3dplot}
\usepackage{subcaption}
\usepackage{caption}
\usepackage{csquotes}
\usepackage{cuted}
\usepackage{Functions_Sphere}

\defcitealias{deutsch_electromagnetic_1955}{D55}
\defcitealias{griffiths_introduction_2017}{G17}
\defcitealias{stratton_electromagnetic_1941}{S41}
\newcommand{\Deutsch}[0]{\citetalias{deutsch_electromagnetic_1955}}
\newcommand{\Deutschspace}[0]{\citetalias{deutsch_electromagnetic_1955} }
\newcommand{\Griffiths}[0]{\citetalias{griffiths_introduction_2017}}
\newcommand{\Griffithsspace}[0]{\citetalias{griffiths_introduction_2017} }
\newcommand{\Stratton}[0]{\citetalias{stratton_electromagnetic_1941}}
\newcommand{\Strattonspace}[0]{\citetalias{stratton_electromagnetic_1941} }

\usepackage{hyperref}
\hypersetup{
    colorlinks=true,
    linkcolor=blue,
    citecolor=blue, 
    filecolor=magenta,      
    urlcolor=blue,
    pdftitle={VRD\_Model\_Derivations},
    }
\newcommand{\vect}[1]{\bm{#1}}
\newcommand{\norm}[1]{\left\lVert#1\right\rVert}
\newcommand{\oddeven}[2]{\genfrac{}{}{0pt}{}{#1}{#2}}
\newcommand{\unitvect}[1]{\,\mathbf{\hat{#1}}}
\newcommand{\angunitvect}[1]{\,\bm{\hat{#1}}}

\tikzset{viewport/.style 2 args={
    x={({cos(-#1)*\RadiusSphere cm},{sin(-#1)*sin(#2)*\RadiusSphere cm})},
    y={({-sin(-#1)*\RadiusSphere cm},{cos(-#1)*sin(#2)*\RadiusSphere cm})},
    z={(0,{cos(#2)*\RadiusSphere cm})}
}}

\title{A Pedagogical Review of the Vacuum Retarded Dipole Model of Pulsar Spin Down}

\author{J. C. Satherley}
\affiliation{School of Physical and Chemical Sciences, University of Canterbury, Christchurch, New Zealand}
\email[J. C. Satherley]{jsa113@uclive.ac.nz}

\author{C. Gordon}
\affiliation{School of Physical and Chemical Sciences, University of Canterbury, Christchurch, New Zealand}

\doi{\%}

\received {dd Mmm YYYY}
\revised  {dd Mmm YYYY}
\accepted {dd Mmm YYYY}
\published{19 March 2021}


\begin{document}

\begin{abstract}
Pulsars are rapidly spinning highly magnetised neutron stars. Their spin period is observed to decrease with time. An early analytical model for this process was the vacuum retarded dipole (VRD) by \cite{deutsch_electromagnetic_1955}. This model assumes an idealised star and it finds that the rotational energy is radiated away by the electromagnetic fields. This model has been superseded by more realistic numerical simulations that account for the non-vacuum like surroundings of the neutron star. However, the VRD still provides a reasonable approximation and is a useful limiting case that can provide some qualitative understanding. We provide detailed derivations of the spin down and related electromagnetic field equations of the VRD solution. We also correct typographical errors  in the general field equations and boundary conditions used by \cite{deutsch_electromagnetic_1955}.
\end{abstract}

\section{INTRODUCTION }

Pulsars are highly magnetised rapidly rotating neutron stars that emit electromagnetic radiation almost across the entire spectrum, from radio to $\gamma$-rays \citep{travelle_pulsars_2011}. Because of the extreme magnetic and electric fields, the environment around a pulsar proves to be interesting and is believed to be the main cause of a pulsar's energy emission. This energy emission results in a decreasing spin-rate for the neutron star. An analytical approximation for this process is given by the Vacuum Retarded Dipole (VRD) model which was originally derived by
\cite{deutsch_electromagnetic_1955} (hereafter referred to as \Deutsch).

The derivations of the results in \Deutschspace are very concise, with many details missing. \textcolor{black}{The purpose of this review is to rederive the main results of \Deutschspace and in doing so provide details on how this derivation takes place mathematically and physically.} We also uncover three typographical errors in \Deutsch, \textcolor{black}{the} most important of which is in our Eqn.~\eqref{eqn:gen_E_theta}, the polar component of the electric field. It is worth noting the first papers we found that mentioned the error in the general field equations are \cite{melatos_spin-down_1997} and \cite{michel_electrodynamics_1999}. However, they do not show the derivations for the general field equations and how the typographical error is found. A reasonably detailed derivation of the VRD solution, without the typos, is shown in 
the textbook by \cite{michel_theory_1991}. However, the method used by Michel is different from \Deutsch. While \Deutschspace uses methods outlined by \cite{stratton_electromagnetic_1941}, Michel just considers multipole expansions at an inclination and does not derive the normalisations. Also the following working carried out by us goes into greater detail for the derivations. 
\textcolor{black}{In this review we use methods outlined in \cite{stratton_electromagnetic_1941} (hereafter referred to as \Stratton). In particular, we refer frequently to Chapters VII and VIII, which consider spherical waves and radiation respectively.} We use \cite{griffiths_introduction_2017} (hereafter referred to as \Griffiths) as a reference for standard electrodynamics theory.

In Sec. \ref{sec:Internal Fields of An Idealised Star}, we begin by outlying the definition of an idealised star, along with fundamental equations that apply due to this assumption, which include Maxwell's equations. We also note down \textcolor{black}{the} Alfvén theorem and a brief description of a vector field. In Sec. \ref{sec:external_fields}, we define the \emph{magnetic field symmetry axis} and its associated coordinate system. The magnetic field is assumed to take the form of a dipole about this axis. The transformation between the two coordinate systems is shown and used to derive the electromagnetic field boundary conditions in the rotation symmetry axis coordinates. In Sec. \ref{sec:general_field_equations}, we show the derivations for the time-dependent and static components of the electromagnetic fields using series expansions and matching coefficients to the boundary conditions. In Sec. \ref{sec:radiation_fields}, we simplify the form of the electromagnetic fields for a large radius, then the power and spin down equations are derived.

\section{The Internal Fields of An Idealised Star}
\label{sec:Internal Fields of An Idealised Star}
For the VRD solution, we consider the electric and magnetic fields of an \textit{idealised star}. An idealised star is a star that is a sharply bounded sphere, is isotropic, is perfectly conducting, and has permittivity and permeability of a vacuum everywhere inside.
The boundary is not continuous but rather an immediate change between the star's medium and the surrounding vacuum. Which, in the case of a real star is not true. However, it greatly simplifies the situation.  

\subsection{Maxwell's Equation}
 The Maxwell equations in SI units are,
\begin{gather}
    \nabla\cdot\vect{E}=\frac{\rho}{\epsilon_0}, \label{eqn:gauss_elec}\\
    \nabla\cdot\vect{H}=0, \label{eqn:gauss_mag}\\
    \nabla\times\vect{E}=-\mu_0\frac{\partial\vect{H}}{\partial t}, \label{eqn:faradays}\\
    \nabla\times\vect{H}=\vect{J}+\epsilon_0\frac{\partial\vect{E}}{\partial t}, \label{eqn:amperes}
\end{gather}
where $\vect{E}$ is the electric field, $\vect{H}$ is the magnetic field (related to $\vect{B}$ by $\vect{B}=\mu_0\vect{H}$),
$\epsilon_0$ is the permittivity of free space, and $\mu_0$
the permeability of free space. $\vect{H}$ will typically be referred to as the $H$ field. In \Deutsch, there is a missing minus sign in his equivalent of equation (\ref{eqn:faradays}). This is clearly a typo as he takes into account the minus sign in later parts of his article.

\subsection{The H field}
Firstly, \Deutschspace describes the typical spherical polar coordinates around the star's axis of rotation, $(r,\theta,\varphi)$, with rotation axis $\vect{\omega}$ aligned with the $z$-axis and origin at the centre of the star. Where $r$ is the radius, $\theta$ is the angle from the pole and $\varphi$ is the azimuth angle measured counterclockwise from the $x$-axis.

Let $\vect{H}(r,\theta,\varphi,t)$ be an arbitrary solenoidal function within the star. A solenoidal function is one for which the divergence is zero at all points in the field ($\nabla\cdot\vect{H}=0$). \Deutschspace states that a necessary and sufficient condition that $\vect{H}$ be ``frozen into'' the star is that
\begin{equation}
\label{eqn:mag_field_int_theta}
\resizebox{0.91\hsize}{!}{%
    $\vect{H}(r,\theta,\varphi,t)=r_0H_r(r,\theta,\lambda)+\theta_0H_\theta(r,\theta,\lambda)+\varphi_0H_\varphi(r,\theta,\lambda),$
    }
\end{equation}
where $\lambda$ is an azimuthal coordinate measured from a meridian fixed in the star\textcolor{black}{, and $r_0$, $\theta_0$ and $\phi_0$ are the unit vectors in the respective coordinate directions}. That is the time and azimuthal angle $\varphi$ can be replaced by $\lambda$ as the $H$ field and the meridian move with the star. The variable $\lambda$ is defined more clearly in eqn.~\eqref{eqn:lambda} and Fig.~\ref{fig:omega_and_e}. $H_r$, $H_\theta$ and $H_\varphi$ give the magnitude of the field in each basis vector's direction. The concept of the magnetic field being ``frozen into'' a medium comes from \textit{Alfvén's Theorem} \citep{roberts_alfvens_2007}.

\subsubsection{Vector Fields}
As $\vect{H}$ is a vector field, at a particular point $(r,\theta,\varphi)$ in space, the field can be represented as a vector with some magnitude in each unit vector's direction. In \Deutsch, the directions are given by the subscript $0$ terms, and the magnitudes are given by the functions proceeding the unit vector terms. In polar coordinates, $r_0$ is the radial component, pointing away from the centre of the sphere. The polar component is $\theta_0$, its direction is tangent to the polar angle. Lastly, $\varphi_0$ is the azimuthal component which has a direction tangent to the azimuthal angle. This notation continues throughout D55. However, we will here on forth elect to use $\unitvect{r}$, $\angunitvect{\theta}$ and $\angunitvect{\varphi}$ to represent these components of the vector field. A visual representation of the unit vectors of a field in polar coordinates is shown in Fig.~\ref{fig:polar_vector_field}. Hence, (\ref{eqn:mag_field_int_theta}) in the notation used for the rest of the report becomes,
\begin{equation*}
    \vect{H}(r,\theta,\varphi,t)=H_r(r,\theta,\lambda)\unitvect{r}+H_\theta(r,\theta,\lambda)\angunitvect{\theta}+H_\varphi(r,\theta,\lambda)\angunitvect{\varphi}.
\end{equation*}

\def\Rotation{160}
\def\Tilt{15}
\def\RadiusSphere{2}
\def\VectorLen{1.5}

\def\VectorFieldPolar{45}
\def\VectorFieldAzimuth{45}

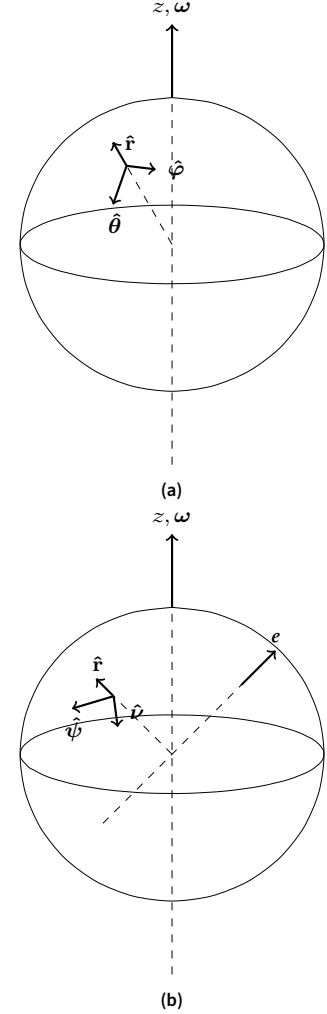
\begin{figure}[ht]
    \begin{subfigure}{0.5\textwidth}
        \centering
        \begin{tikzpicture}
            \begin{scope}[viewport={\Rotation}{15}, very thin]
            
                \draw[dashed,-] (0,0,-\VectorLen) -- (0,0,1);
                \draw[thick,->] (0,0,1) -- (0,0,\VectorLen) node[anchor=south]{$z, \vect{\omega}$};
                
                \draw[domain=0-\Rotation:360-\Rotation, variable=\azimuth, smooth] plot (\ToXYZ{90}{\azimuth});
                \draw[domain=0:360, variable=\elevation, smooth] plot (\ToXYZ{\elevation}{\Rotation});
                
                \draw[domain=0:1, variable=\radius, dashed] plot (\ToXYZr{\radius}{\VectorFieldPolar}{\VectorFieldAzimuth});
                \draw[domain=1:1.3, variable=\radius, smooth, thick, ->] plot (\ToXYZr{\radius}{\VectorFieldPolar}{\VectorFieldAzimuth}) node[right]{$\unitvect{r}$};
                \draw[domain=0:0.3, variable=\t, thick, ->] plot (\ThetaTangentGradient{1}{\VectorFieldPolar}{\VectorFieldAzimuth}{\t}) node[anchor=north]{$\angunitvect{\theta}$};
                \draw[domain=0:0.3, variable=\t, thick, ->] plot (\PhiTangentGradient{1}{\VectorFieldPolar}{\VectorFieldAzimuth}{\t}) node[anchor=west]{$\angunitvect{\varphi}$};
            \end{scope}
        \end{tikzpicture}               
        \caption{}
        \label{fig:polar_vector_field}
    \end{subfigure}
    \begin{subfigure}{0.5\textwidth}
        \centering
        \begin{tikzpicture}
            \begin{scope}[viewport={\Rotation}{\Tilt}, very thin]
            
                \draw[thick,->] (0,0,1) -- (0,0,\VectorLen) node[anchor=south]{$z, \vect{\omega}$};
                \draw[dashed,-] (0,0,-\VectorLen) -- (0,0,1);
                
                \draw[domain=0:360, variable=\azimuth, smooth] plot (\ToXYZ{90}{\azimuth});
                \draw[domain=0:360, variable=\elevation, smooth] plot (\ToXYZ{\elevation}{\Rotation});
                
            \end{scope}
            \tdplotsetmaincoords{0}{0}
            \tdplotsetrotatedcoords{60-\Tilt}{40}{-90-\Tilt}
            
            \def\EVectorFieldPolar{50}
            \def\EVectorFieldAzimuth{240}
            
            \begin{scope}[tdplot_rotated_coords, very thin]
                \draw[domain=-2:2, variable=\radius, dashed] plot (\ToXYZr{\radius}{0}{0});
                \draw[domain=2:3, variable=\radius, thick, ->] plot (\ToXYZr{\radius}{0}{0}) node[anchor=south]{$\vect{e}$};
            
                \draw[domain=0:2, variable=\radius, dashed] plot (\ToXYZr{\radius}{\EVectorFieldPolar}{\EVectorFieldAzimuth});
                \draw[domain=2:2.6, variable=\radius, smooth, thick, ->] plot (\ToXYZr{\radius}{\EVectorFieldPolar}{\EVectorFieldAzimuth}) node[above]{$\unitvect{r}$};
                \draw[domain=0:0.3, variable=\t, thick, ->] plot (\ThetaTangentGradient{2}{\EVectorFieldPolar}{\EVectorFieldAzimuth}{\t}) node[anchor=north]{$\angunitvect{\psi}$};
                \draw[domain=0:0.3, variable=\t, thick, ->] plot (\PhiTangentGradient{2}{\EVectorFieldPolar}{\EVectorFieldAzimuth}{\t}) node[above right]{$\angunitvect{\nu}$};
                
            \end{scope}
        \end{tikzpicture}               
        \caption{}
        \label{fig:Epolar_vector_field}
    \end{subfigure}
    \caption{(a) The unit vectors of a vector field at a point on a sphere in $(r, \theta, \varphi)$. (b) The unit vectors of a vector field at a point on a sphere in $(r,\psi,\nu)$.}
\end{figure}

\section{The External Field in Vacuum}
\label{sec:external_fields}
It is convenient to define a new set of polar coordinates $(r,\psi,\nu)$ with axis $\vect{e}$ such that the magnetic field is symmetrical about $\vect{e}$ (known as the \emph{magnetic field symmetry axis}). The axis $\vect{e}$ will be in fixed rotation with the star due to Alfvén's theorem. This new coordinate system is shown in Fig.~\ref{fig:e_coords}. The magnetic field symmetry here refers to, for a fixed $r$ and $\psi$, the magnitude of the field does not change as $\nu$ is changed and the direction does not change relative to the local unit vectors. The angle between the rotation axis $\vect{\omega}$ and $\vect{e}$ is $\chi$. This relationship is shown in Fig.~\ref{fig:omega_and_e}. The axis $\vect{e}$ has polar angle $\chi$ and azimuthal angle $\omega t$ in $(r,\theta,\varphi)$, $\vect{e}=(r,\chi,\omega t)$, at some time $t$ during the rotation of the star. The azimuthal angle $\lambda$ mentioned earlier is measured from $\vect{e}$ along $\varphi$, so that 
\begin{equation}
\label{eqn:lambda}
    \lambda=\varphi-\omega t. 
\end{equation}

\def\Rotation{160}
\def\Tilt{15}
\def\RadiusSphere{2}
\def\VectorLen{1.5}

\def\VectorFieldPolar{45}
\def\VectorFieldAzimuth{45}

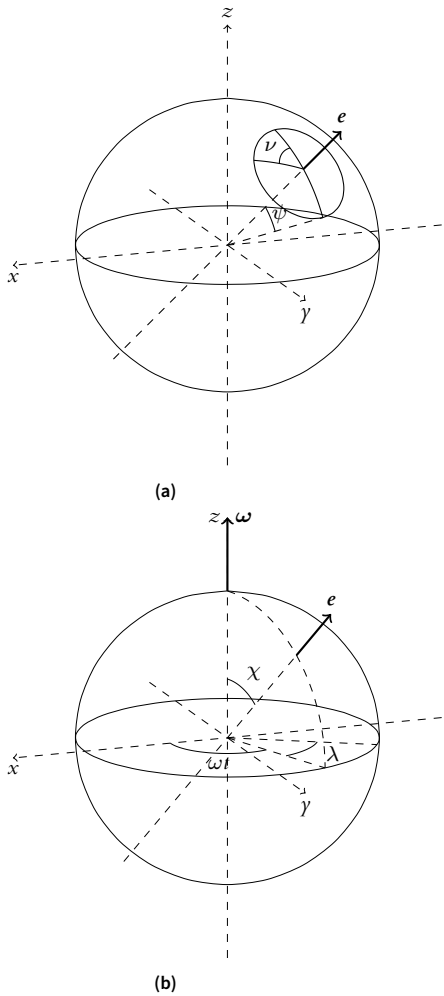
\begin{figure}[ht]
    \begin{subfigure}{0.5\textwidth}
    \centering
        \begin{tikzpicture}
            \begin{scope}[viewport={160}{15}, very thin]
            
                \draw[dashed,->] (\ToXYZr{-1.5}{90}{0}) -- (\ToXYZr{1.5}{90}{0}) node[anchor=north]{$x$};
                \draw[dashed,->] (\ToXYZr{-1.5}{90}{90}) -- (\ToXYZr{1.5}{90}{90}) node[anchor=north]{$y$};
                \draw[dashed,->] (\ToXYZr{-1.5}{0}{0}) -- (\ToXYZr{1.5}{0}{0}) node[anchor=south]{$z$};
                
                \draw[domain=0-\Rotation:360-\Rotation, variable=\azimuth, smooth] plot (\ToXYZ{90}{\azimuth});
                \draw[domain=0:360, variable=\elevation, smooth] plot (\ToXYZ{\elevation}{\Rotation});
            \end{scope}
            \tdplotsetmaincoords{0}{0}
            \tdplotsetrotatedcoords{45}{45}{-112.5}
            \begin{scope}[tdplot_rotated_coords]
                \draw[domain=-3:2, variable=\radius, dashed] plot (\ToXYZr{\radius}{0}{0});
                \draw[domain=2:3, variable=\radius, thick, ->] plot (\ToXYZr{\radius}{0}{0}) node[anchor=south]{$\vect{e}$};
                
                \draw[domain=0:360, variable=\azimuth, smooth] plot (\ToXYZr{2}{20}{\azimuth});
                \draw[dashed] (\ToXYZr{0}{20}{0}) -- (\ToXYZr{2}{20}{0});
                \draw[domain=0:20, variable=\polar, smooth] plot (\ToXYZr{1}{\polar}{0}) node[] at (\ToXYZr{1.2}{10}{0}) {$\psi$};
                \draw[domain=-20:20, variable=\polar, smooth] plot (\ToXYZr{2}{\polar}{0});
                \draw[domain=180:240, variable=\azimuth, smooth] plot (\ToXYZr{2}{10}{\azimuth}) node[] at (\ToXYZr{2}{15}{210}) {$\nu$};
                \draw[domain=0:20, variable=\polar, smooth] plot (\ToXYZr{2}{\polar}{240});
            \end{scope}
        \end{tikzpicture}
        \caption{}
        \label{fig:e_coords}
    \end{subfigure}
    \begin{subfigure}{0.5\textwidth}
    \centering
        \begin{tikzpicture}
            \begin{scope}[viewport={\Rotation}{15}, very thin]
            
                \draw[dashed,->] (-\VectorLen,0,0) -- (\VectorLen,0,0) node[anchor=north]{$x$};
                \draw[dashed,->] (0,-\VectorLen,0) -- (0,\VectorLen,0) node[anchor=north]{$y$};
                \draw[dashed,->] (0,0,-\VectorLen) -- (0,0,\VectorLen) node[anchor=east]{$z$};
                
                \draw[thick,->] (0,0,1) -- (0,0,\VectorLen) node[anchor=west]{$\vect{\omega}$};
                
                \draw[domain=0-\Rotation:360-\Rotation, variable=\azimuth, smooth] plot (\ToXYZ{90}{\azimuth});
                \draw[domain=0:360, variable=\elevation, smooth] plot (\ToXYZ{\elevation}{\Rotation});
                
                \def\Epolar{45}
                \def\Eazimuth{110}
                
                \draw[domain=-1.5:1, variable=\radius, dashed] plot (\ToXYZr{\radius}{\Epolar}{\Eazimuth});
                \draw[domain=1:\VectorLen, variable=\radius, thick, ->] plot (\ToXYZr{\radius}{\Epolar}{\Eazimuth}) node[anchor=south]{$\vect{e}$};
                
                \draw[domain=0:\Epolar, variable=\elevation, smooth] plot (\ToXYZr{0.4}{\elevation}{\Eazimuth}) node[] at (\ToXYZr{0.5}{22.5}{135}) {$\chi$};
                \draw[domain=0:\Eazimuth, variable=\azimuth, smooth] plot (\ToXYZr{0.4}{90}{\azimuth}) node[] at (\ToXYZr{0.65}{90}{\Eazimuth/2+10}) {$\omega t$};
                \draw[domain=0:90, variable=\elevation, dashed] plot (\ToXYZr{1}{\elevation}{\Eazimuth});
                \draw[domain=0:1, variable=\radius, dashed] plot (\ToXYZr{\radius}{90}{\Eazimuth});
                
                \draw[domain=\Eazimuth:\Eazimuth+40, variable=\azimuth, smooth] plot (\ToXYZr{0.6}{90}{\azimuth}) node[] at (\ToXYZr{0.8}{90}{\Eazimuth+40/2}) {$\lambda$};
                \draw[domain=0:1, variable=\radius, dashed] plot (\ToXYZr{\radius}{90}{\Eazimuth+40});
            \end{scope}
        \end{tikzpicture}
        \caption{}
        \label{fig:omega_and_e}
    \end{subfigure}
    \caption{The relation between the two coordinate systems $(r,\theta,\varphi)$ and $(r,\psi,\nu)$. $x$, $y$, and $z$ are the Cartesian coordinates. (a) The coordinate system $(r,\psi,\nu)$ portrayed on a sphere. (b) The star's rotation axis $\vect{\omega}$ and the magnetic field's symmetry axis $\vect{e}$. $\chi$ is the angle of inclination between $\vect{\omega}$ and $\vect{e}$, and $\lambda$ is measured from the projection of $\vect{e}$ onto the xy-plane.}
    
\end{figure}

We shall now consider a magnetic field which within the star has the form,
\begin{equation}
\label{eqn:mag_field_inside_star}
    \vect{H}=R_1(r)S_1(\psi)\unitvect{r}+R_2(r)S_2(\psi)\angunitvect{\psi}+R_3(r)S_3(\psi)\angunitvect{\nu},
\end{equation}
where $R_i$ and $S_i$ are arbitrary functions of $r$ and $\psi$ respectively, with $i=1,2$, or $3$. Note that $\vect{H}$ has no dependence on $\nu$ for the magnitude of the vectors and hence the magnitude is symmetrical about axis $\vect{e}$ as stated earlier. In this situation \Deutschspace assumes the interior magnetic field to be a \emph{dipole} about $\vect{e}$ by taking $S_1(\psi)=\cos\psi$.

\subsection{Matrix Transformation between Vector Field Basis Vectors}
A vector transformation between the two coordinate systems mentioned is required whilst working out the boundary conditions for the fields in both coordinate systems. It is convenient to figure out the transformation now, then approach the boundary conditions. Define the transformation matrix $R:(\unitvect{r},\angunitvect{\theta},\angunitvect{\varphi})\to(\unitvect{r},\angunitvect{\psi},\angunitvect{\nu})$ which rotates between the two unit vector systems at a point in space. 
Figs. \ref{fig:polar_vector_field} and \ref{fig:Epolar_vector_field} show the two sets of basis vectors at the same point. It is clear that $\unitvect{r}$ points radially out, normal to the sphere's surface, in both figures. Hence, at the same point in space $\unitvect{r}$ is the same in both coordinate systems. However, the angular components undergo a rotation as $\angunitvect{\theta}, \angunitvect{\varphi}, \angunitvect{\psi}$ and $\angunitvect{\nu}$ all lie in a plane tangent to the sphere. The rotation between these unit vectors will be a two dimensional rotation about $\unitvect{r}$. Let the angle of this rotation be $\xi$, the transformation matrix $R$ is then
\begin{equation}
\label{eqn:rotation_matrix}
    R=
    \begin{bmatrix}
    1 & 0 & 0\\
    0 & \cos\xi & -\sin\xi \\
    0 & \sin\xi & \cos\xi
    \end{bmatrix}.
\end{equation}
This matrix transformation takes a basis of vectors at some point in space from $(\unitvect{r},\angunitvect{\theta},\angunitvect{\varphi})$ and rotates them anticlockwise by an angle $\xi$ with $\unitvect{r}$ as axis to the unit vectors $(\unitvect{r},\angunitvect{\psi},\angunitvect{\nu})$. To \textit{reverse} this transformation the angle of rotation is replaced by $-\xi$. 

The angle $\xi$ can be found in terms of already defined coordinate parameters and values. This is done by considering triangles on the surface of a sphere. The spherical triangle for this situation can be found in Fig.~\ref{fig:Spherical Triangle} in \ref{Appendix:SphericalTrangle}. Then from solutions of spherical triangles (\ref{eqn:sphi_tri_sin}) and (\ref{eqn:sph_tri_3_sides}) the following relationships can be found,
\begin{equation}
\label{eqn:xi_defined}
    \begin{aligned}
    \sin\xi&=\frac{\sin\chi\sin\lambda}{\sin\psi},\\
    \cos\xi&=\frac{\cos\chi-\cos\psi\cos\theta}{\sin\psi\sin\theta}.
    \end{aligned}
\end{equation}
An interesting relationship that is obtained from the standard spherical triangle relation (\ref{eqn:sph_tri_2_sides}) is
\begin{equation}
\label{eqn:cos_psi}
    \cos\psi=\cos\chi\cos\theta+\sin\chi\sin\theta\cos\lambda,
\end{equation}
which will be required momentarily.

\subsection{Boundary Conditions of the Fields}
As the surface of the star is a rigid boundary between the interior of the star and free space, the necessary components of the external electromagnetic fields must satisfy the boundary conditions at $r=a$ (see \Griffiths). These boundary conditions are set by the internal electromagnetic fields. From equations (7.61) and (7.62) of \Griffiths,
the magnetic fields orthogonal to the surface must be equal and the electric fields tangent to the surface must be equal. Hence, using (\ref{eqn:mag_field_inside_star}), the first condition at the surface of the star that must be satisfied is,
\begin{equation*}
    H_r^{\text{ext}}=H_r^{\text{int}}=R_1(a)S_1(\psi).
\end{equation*}


Next we look at the internal electric field.
To do this we use Ohm's law from \Griffiths, Eqn.~7.2, that relates the electric field and current
density, $\vect{J}$ and is given by,
\begin{equation}
\vect{E}+\mu_0\vect{V}\times\vect{H}=\frac{\vect{J}}{\sigma}\label{eq:ohm}
\end{equation}
where $\sigma$ is the conductivity and $\vect{V}$ is the local velocity of rotation of the star. Assuming $\sigma\rightarrow \infty$ gives
\begin{equation}
 \label{eqn:int_ele_field}
     \vect{E}=-\mu_0\vect{V}\times\vect{H}.
 \end{equation}
The velocity in $(\unitvect{r},\angunitvect{\theta},\angunitvect{\varphi})$ is
\begin{equation}
\label{eqn:vel_of_star}
    \vect{V}=\vect{\omega}\times\vect{r}=r\omega\sin\theta\angunitvect{\varphi}.
\end{equation}

However, the vector field (\ref{eqn:mag_field_inside_star}) must be converted from $(\unitvect{r},\angunitvect{\psi},\angunitvect{\nu})$ to $(\unitvect{r},\angunitvect{\theta},\angunitvect{\varphi})$ so that $\vect{V}$ and $\vect{H}$ are defined on the same basis. This is done by rotating (\ref{eqn:mag_field_inside_star}) using (\ref{eqn:rotation_matrix}). However, as the reverse transformation is needed, 
the signs of $R_{2,3}$ and $R_{3,2}$ elements are switched. 
\begin{equation*}
    \begin{aligned}
    R(-\xi)\vect{H}=&R_1(r)S_1(\psi)\unitvect{r}\\
    &+\left[R_2(r)S_2(\psi) \cos\xi + R_3(r)S_3(\psi) \sin\xi \right]\angunitvect{\theta}\\
    &+\left[-R_2(r)S_2(\psi) \sin\xi + R_3(r)S_3(\psi) \cos\xi \right]\angunitvect{\varphi}.
    \end{aligned}
\end{equation*}
Applying the previous result to (\ref{eqn:int_ele_field}) with \eqref{eqn:vel_of_star} and using Fig.~\ref{fig:Spherical Triangle} of \ref{Appendix:SphericalTrangle} we can find the $E$ field inside the star,
\begin{equation*}
    \begin{aligned}
    \vect{E}=&-\mu_0\vect{V}\times(R(-\xi)\vect{H})\\
    =&\mu_0 r\omega\sin\theta\left(R_2(r)S_2(\psi) \cos\xi + R_3(r)S_3(\psi) \sin\xi \right)\unitvect{r}\\
    &-\omega\mu_0rR_1(r)S_1(\psi)\sin\theta\angunitvect{\theta}\\
    &+0\angunitvect{\varphi},
    \end{aligned}
\end{equation*}
and evaluating this at $r=a$ gives the tangential components as
\begin{equation*}
    \begin{aligned}
    E_\theta&=-\omega\mu_0aR_1(a)S_1(\psi)\sin\theta,\\
    E_\varphi&=0.
    \end{aligned}
\end{equation*}

All but the third boundary condition match what \Deutschspace obtained. There is a typographical error in \Deutsch's boundary conditions (Deutsch's Eqn.~(10)), the third condition should be written as $E_\varphi^{\text{ext}}=E_\varphi^{\text{int}}=0$ rather than $E_r^{\text{ext}}=E_\varphi^{\text{int}}=0$. 

A more useful form of the boundary conditions can be found. In particular, the field equations are solved for when $S_1=\cos\psi$. So by recalling (\ref{eqn:cos_psi}), the boundary conditions then results in,
\begin{equation}
\label{eqn:boundary_conditions_extended}
\resizebox{\linewidth}{!}{$
    \begin{aligned}
    H_r^{\text{ext}}=H_r^{\text{int}}&=R_1(a)(\cos\chi\cos\theta+\sin\chi\sin\theta\cos\lambda),\\
    E_\theta^{\text{ext}}=E_\theta^{\text{int}}&=-\omega\mu_0aR_1(a)(\cos\chi\cos\theta+\sin\chi\sin\theta\cos\lambda)\sin\theta\\
    &=-\frac{1}{2}\omega\mu_0aR_1(a)(\sin{2\theta}\cos\chi+(1-\cos{2\theta})\sin\chi\cos\lambda),\\
    E_\varphi^{\text{ext}}=E_\varphi^{\text{int}}&=0.
    \end{aligned}
    $}
\end{equation}
From the boundary conditions in the $(r,\theta,\phi)$ coordinate system, there is a static (time-independent) component and a time-dependent component for the fields. The terms in \eqref{eqn:boundary_conditions_extended} which relate to the time-dependent component have a dependency on $\lambda$, which depends on time from \eqref{eqn:lambda}.

\section{Derivation of General Field Equations}
\label{sec:general_field_equations}
The general field equations for the VRD solution external to the star are provided in the Appendix of \Deutsch. However, with minimal explanation given in \Deutschspace about the derivation of the field equations, we wish to show how they are found. 
We use results found in Chapters VII and VIII of \Strattonspace for this section. \textcolor{black}{Chapter VII proves beneficial in supplying general solutions to the vector wave equation which, must be satisfied by the time evolving electromagnetic fields around the neutron star. We apply these solutions to the neutron star situation by firstly considering} the \textit{propagation factor} of the waves outside the star. In \Stratton, on page 392, the propagation factor (complex wavenumber) is given as, 
\begin{equation}
    k^2=\epsilon\mu\omega^2+i\sigma\mu\omega,
\end{equation}
and as we are now considering the fields external of the star, in a vacuum, $\epsilon=\epsilon_0$, $\mu=\mu_0$, $\sigma=0$ and $c=1/\sqrt{\mu_0\epsilon_0}$. Hence,
\begin{equation}
\label{eqn:wavenumber}
    k = \omega/c.
\end{equation}
Then for convenience define,
\begin{equation}
\label{eqn:rho_def}
    \begin{gathered}
    \rho=kr=(\omega/c)r,\\
    \alpha=ka=(\omega/c)a,
    \end{gathered}
\end{equation}
where $\omega$ is the rotational velocity of the star, $c$ the speed of light, and $a$ is the radius of the star. Here $\rho$ and $\alpha$ are unitless numbers.
A trivial relationship that will be required is
\begin{equation*}
    \frac{\partial}{\partial\rho}=\frac{\omega}{c}\frac{\partial}{\partial r}.
\end{equation*}

Next consider the EM fields surrounding the star in two parts, a static time-independent component and a time-dependent component, where the complete field will be the linear combination of the two components (\Griffiths). \textcolor{black}{The time-independent component will be axially symmetric about the axis $\vect{\omega}$ and hence have no $\varphi$ dependence ($m=0$, $m$ will be defined clearer shortly). Whilst the time-dependent component will be non-axially symmetric and hence have a dependence on $\varphi$ ($m\neq0$).} This can be expressed as
\begin{equation*}
\begin{gathered}
    \vect{H} = \vect{H}_{\text{static}}(r,\theta)+\vect{H}_{\text{time-dependent}}(r,\theta, \lambda),\\
    \vect{E} = \vect{E}_{\text{static}}(r,\theta)+\vect{E}_{\text{time-dependent}}(r,\theta, \lambda).
\end{gathered}
\end{equation*}

We first consider the time-dependent component in the following section on the wave solution. Then use a multipole expansion to provide the static component.

\subsection{Wave Solution}
Due to the rotating nature of the magnetic field symmetry axis around the star's axis of rotation, the vector fields $\vect{E}$ and $\vect{H}$ satisfy the same vector differential wave equation from \Stratton, Sec. 7.1, Eqn.~(1), which is given as
\begin{equation}
\label{eqn:wave_eqn}
    \nabla^2\vect{C}-\mu\epsilon\frac{\partial^2\vect{C}}{\partial t^2}-\mu\sigma\frac{\partial \vect{C}}{\partial t}=0,
\end{equation}
where $\vect{C}$ represents either vector $\vect{E}$ or $\vect{H}$. \textcolor{black}{We can express $\vect{E}$ and $\vect{H}$ as a series of solutions $\vect{\mathrm{M}}_n$ and $\vect{\mathrm{N}}_n$ which satisfy the vector differential wave equation \eqref{eqn:wave_eqn} for $\vect{C}$. From \Stratton, Sec. 7.1, Eqn.~(12) this representation is,}
\begin{equation}
\label{eqn:E_H_series}
    \begin{gathered}
        \vect{E}=-\sum_n\left(a_n\vect{\mathrm{M}}_n+b_n\vect{\mathrm{N}}_n\right),\\
        \vect{H}=-\frac{k}{i\omega\mu_0}\sum_n\left(a_n\vect{\mathrm{N}}_n+b_n\vect{\mathrm{M}}_n\right)
    \end{gathered}
\end{equation}
\textcolor{black}{where $\vect{\mathrm{M}}_n$ and $\vect{\mathrm{N}}_n$ have more recently been referred to as the Magnetic and Electric harmonics respectfully. In particular, the functions are expressed as $\vect{\mathrm{M}}_n=\nabla\times\vect{a}\psi$ and $\vect{\mathrm{N}}_n=(1/k)\nabla\times\vect{\mathrm{M}}_n$, where $\vect{a}$ is a constant vector and $\psi$ is a scalar function satisfying $\nabla^2\psi+k^2\psi=0$, which is known as the Helmholtz equation. The derivation for the general solutions of $\vect{\mathrm{M}}_n$ and $\vect{\mathrm{N}}_n$ is provided in \Strattonspace and the results are quoted below.} 
These solutions will also contain a time dependence. Due to the linearity of the wave equation, the time dependency can be split off such that $\vect{\mathrm{M}}=\vect{\mathrm{m}}e^{-i\omega t}$ and $\vect{\mathrm{N}}=\vect{\mathrm{n}}e^{-i\omega t}$ without any loss of generality. Then from \Stratton, Sec. 7.11, Eqns. (11) and (12), and rewriting them in a slightly more useful form with $\rho$ instead of $kr$, 
\begin{equation}
\label{eqn:m_wave_vec}
    \begin{aligned}
    \vect{\mathrm{m}}_{\oddeven{e}{o} {m\,n}}=&\mp \frac{m}{\sin \theta}h^{(1)}_{n}(\rho) P_{n}^{m}(\cos \theta) \oddeven{\sin}{\cos}(m \varphi)\angunitvect{\theta}\\
    &-h^{(1)}_{n}(\rho)\frac{\partial P_{n}^{m}}{\partial \theta}\oddeven{\cos}{\sin}(m\varphi)\angunitvect{\varphi},
    \end{aligned}
\end{equation}
\begin{equation}
\label{eqn:n_wave_vec}
    \begin{aligned}
    \vect{\mathrm{n}}_{\oddeven{e}{o} {m\,n}}=&\frac{n(n+1)}{\rho} h^{(1)}_{n}(\rho) P_{n}^{m}(\cos \theta) \oddeven{\cos }{\sin }(m \varphi)\unitvect{r}\\
    &+\frac{1}{\rho} \frac{\partial}{\partial \rho}\left[\rho\,h^{(1)}_{n}(\rho)\right] \frac{\partial}{\partial \theta} P_{n}^{m}(\cos \theta) \oddeven{\cos}{\sin}(m \varphi)\angunitvect{\theta}\\
    &\mp \frac{m}{\rho \sin \theta} \frac{\partial}{\partial \rho}\left[\rho\, h^{(1)}_{n}(\rho)\right] P_{n}^{m}(\cos \theta)\oddeven{\sin}{\cos}(m \varphi)\angunitvect{\varphi},
    \end{aligned}
\end{equation}
where $h^{(1)}_n$ is the spherical Bessel function of the third kind (refer to \ref{appendix:bessel_func}),  $P_{n}^{m}(\cos \theta)$ are the associated Legendre polynomials (refer to \ref{appendix:leg_fun})
\textcolor{black}{, and $m$ and $n$ are associated to a particular solution to the Helmholtz equation. In particular, $m$ defines the azimuthal dependence of the $E$ and $H$ fields. Whilst $n$ controls the polar dependence. These functions \eqref{eqn:m_wave_vec} and \eqref{eqn:n_wave_vec} are now known as vector spherical harmonics. The subscript $e$ or $o$ define the even and odd solution, this chooses the sign of some terms and whether the trigonometric function for $\varphi$ is $\sin$ or $\cos$.} 
From here on $h^{(1)}_n$ will be referred to as $h_n$. However, \Deutschspace does not use the even-odd notation from \Strattonspace but rather \Deutschspace multiplies the function by a complex exponential then takes the real part of the result to give the wave solutions. This can be expressed as an operator that first multiplies a function $f(x)$ by $e^{im\varphi}$, then takes the real part of the result,
\begin{equation}
\label{eqn:oddeven}
\begin{aligned}
    f(x)\oddeven{\cos}{\sin}(m \varphi)&=\Re(f(x)e^{im\varphi}),\\
    \mp f(x)\oddeven{\sin}{\cos}(m \varphi)&=-\Re(f(x)ie^{im\varphi}),
\end{aligned}
\end{equation}
where $f(x)$ may be complex-valued function and $\Re$ denotes the operator which returns the real part of an expression. The operation \eqref{eqn:oddeven} essentially selects $\cos(m\varphi)$ or $\sin(m\varphi)$ depending on the complex-valued nature of $f(x)$.

We wish to solve for the field equations in polar coordinates around rotation axis $\vect{\omega}$. \textcolor{black}{Due to the dependence of $\varphi$ in our electromagnetic boundary conditions \eqref{eqn:boundary_conditions_extended} only being $\cos(\varphi)$ and $\sin(\varphi)$, we are limited to $m=1$ for our equations \eqref{eqn:m_wave_vec} and \eqref{eqn:n_wave_vec}. Otherwise terms which contain $\cos(m\varphi)$ and $\sin(m\varphi)$, where $m\neq1$, will appear in the solutions of \eqref{eqn:E_H_series} and cause disagreement of the external and internal electromagnetic fields at the surface of the star.}
By also comparing the series expression (\ref{eqn:E_H_series}) and the boundary conditions (\ref{eqn:boundary_conditions_extended}) we see that if a trigonometric function of $\theta$ does not appear in the boundary conditions, its weighting coefficient will be zero (as will be noticed in the derivations shortly). Hence, it is apparent from inspection of the boundary conditions and the Associated Legendre Polynomials that only $n=1$ and $n=2$ are required. So evaluating (\ref{eqn:m_wave_vec}) and (\ref{eqn:n_wave_vec}) for $m=1$, $n=1$ and $n=2$ and including the new notation from \eqref{eqn:oddeven} gives,   
\begin{equation}
\label{eqn:m_n_eval}
    \begin{aligned}
    \vect{\mathrm{m}}_{1,1} =& i h_1(\rho)e^{i\varphi}\angunitvect{\theta} - h_1(\rho)\cos\theta e^{i\varphi}\angunitvect{\varphi},\\
    \vect{\mathrm{m}}_{1,2} =& 3 i h_2(\rho)\cos\theta e^{i\varphi}\angunitvect{\theta} - 3h_2(\rho)\cos{2\theta}e^{i\varphi}\angunitvect{\varphi},\\
    \vect{\mathrm{n}}_{1,1} =& 2\frac{h_1(\rho)}{\rho}\sin\theta e^{i\varphi}\unitvect{r} + \frac{1}{\rho}(\rho h_1'(\rho)+h_1(\rho))\cos\theta e^{i\varphi}\angunitvect{\theta}\\
    & + \frac{i}{\rho}(\rho h_1'(\rho)+h_1(\rho))e^{i\varphi}\angunitvect{\varphi},\\
    \vect{\mathrm{n}}_{1,2}=& 9\frac{h_2(\rho)}{\rho}\sin{2\theta} e^{i\varphi}\unitvect{r} + \frac{3}{\rho}(\rho h_2'(\rho)+h_2(\rho))\cos{2\theta} e^{i\varphi}\angunitvect{\theta}\\
    & + \frac{3i}{\rho}(\rho h_2'(\rho)+h_2(\rho))\cos\theta e^{i\varphi}\angunitvect{\varphi},
    \end{aligned}
\end{equation}
where we have neglected the operator that extracts the real part of the result. This will be added during the calculation for the coefficients in \eqref{eqn:E_H_series}.

\subsection{Multipole Expansion}
A static field symmetric about the rotation axis can be expressed using a multipole expansion. Upon inspection of the static terms in the boundary conditions (\ref{eqn:boundary_conditions_extended}) and the multipole expansions given in \Griffithsspace it can be seen, with some working, that only a magnetic dipole and electric quadrupole are present in this situation based on the conditions that multipole terms are linearly independent of each other and that the boundary conditions will uniquely determine the fields (electromagnetism uniqueness theorem, see \Griffiths).

\subsubsection{Magnetic Dipole}

We can begin by taking a magnetic dipole and check whether it satisfies the $H$ field boundary condition from \eqref{eqn:boundary_conditions_extended}. From equation~(5.88) of
\Griffiths, a magnetic dipole is given as,
\begin{equation}
\label{eqn:magnetic_dipole}
    \vect{H}_{\text{dip}}(\vect{r})=\frac{m}{4\pi r^3}\left(2\cos\theta\unitvect{r}+\sin\theta\angunitvect{\theta}\right),
\end{equation}
where $m$ is the magnetic dipole moment of the field. Now compare the radial component of \eqref{eqn:magnetic_dipole} to the static term of the radial $H$ field boundary condition from \eqref{eqn:boundary_conditions_extended}.
\begin{equation*}
    \frac{m}{4\pi a^3}2\cos\theta = R_1(a)\cos\chi\cos\theta.
\end{equation*}
We see that the function depending on $\theta$ in the boundary condition is satisfied fully by the dipole term. Hence, a magnetic dipole is the only static term present due to the boundary conditions. This implies the ratio $m/(4\pi)$ will satisfy, 
\begin{equation*}
    \frac{m}{4\pi} = \frac{1}{2}a^3R_1(a)\cos\chi.
\end{equation*}
Therefore, the static component of the magnetic field will be
\begin{equation}
\label{eqn:mag_dipole}
    \vect{H}_{\text{static}}(\vect{r})=\frac{1}{2}\frac{a^3}{ r^3}R_1(a)\cos\chi\left(2\cos\theta\unitvect{r}+\sin\theta\angunitvect{\theta}\right).
\end{equation}

\subsubsection{Electric Quadrupole}
We begin by assuming the solution for the static electric field to be a quadrupole, after which we show that it is the only component required to satisfy the boundary conditions due to the dependency on $\theta$. From \Griffiths, equation (3.65), the quadrupole potential term is
\begin{equation*}
    V_{\text{quad}}(r)=\frac{B_2}{r^3}\frac{1}{2}\left(3\cos^2\theta-1\right).
\end{equation*}
where $B_2$ is the weighting coefficient from the series expansion in \Griffiths.

From equation (2.23) of \Griffiths, the electric field is the negative gradient of the potential. Hence, the polar component of the $E$ field is
\begin{equation*}
    E_{\text{quad},\,\theta}=-\frac{1}{r}\frac{\partial V}{\partial\theta} = \frac{B_2}{r^4}3\cos\theta\sin\theta.
\end{equation*}

The functions depending on the variable $\theta$ match those of the boundary condition \eqref{eqn:boundary_conditions_extended}. Therefore, the boundary condition only requires the static component of the electric field to be a quadrupole. Comparing the above result to \eqref{eqn:boundary_conditions_extended} we can uniquely determine $B_2$ for the situation,
\begin{equation*}
\begin{gathered}
    \frac{B_2}{a^4}3\cos\theta\sin\theta = -\omega\mu_0 a R_1(a)\cos\chi\cos\theta\sin\theta,\\
    \implies B_2 = -\frac{1}{3}\omega\mu_0 a^5 R_1(a) \cos\chi.
\end{gathered}
\end{equation*}
and on applying the gradient function to the potential in spherical coordinates (see \ref{appendix:spherical_calculus}),
\begin{equation*}
    \begin{aligned}
    E_{\text{quad},\,r}&=-\frac{\partial V}{\partial r} =-\frac{1}{2}\omega\mu_0 a\frac{a^4}{r^4}R_1(a)\cos\chi\left(3\cos^2\theta-1\right),\\
    E_{\text{quad},\,\theta}&=-\frac{1}{r}\frac{\partial V}{\partial\theta} = - \omega \mu_0 a R_1(a)\frac{a^4}{r^4}\cos\chi\cos\theta\sin\theta,\\
    E_{\text{quad},\,\varphi}&=-\frac{1}{r\sin\theta}\frac{\partial V}{\partial\varphi}=0.
    \end{aligned}
\end{equation*}
The trigonometric identities $3\cos^2\theta-1=\frac{1}{2}\left(3\cos{2\theta}+1\right)$ and $\cos\theta\sin\theta=\frac{1}{2}\sin 2\theta,$
can be used to write the quadrupole terms in a similar form as \Deutsch,
\begin{equation}
\label{eqn:ele_quadrupole}
\begin{aligned}
    E_{\text{quad},\,r}&=-\frac{1}{4}\omega\mu_0 a\frac{a^4}{r^4}R_1(a)\cos\chi\left(3\cos2\theta+1\right),\\
    E_{\text{quad},\,\theta}&= - \frac{1}{2}\omega \mu_0 a R_1(a)\frac{a^4}{r^4}\cos\chi\sin2\theta,\\
    E_{\text{quad},\,\varphi}&=0.
\end{aligned}
\end{equation}

\subsection{Derivation of General Field Equations}
We are now in a position to derive the equations in the Appendix of \Deutsch. The general field equations will be the sum of the multipoles derivations (\ref{eqn:mag_dipole}) and (\ref{eqn:ele_quadrupole}), and the series expressions (\ref{eqn:E_H_series}),
\begin{equation}
\label{eqn:series_expansion}
\resizebox{.89\hsize}{!}{$
    \begin{gathered}
    \vect{H}=\vect{H}_{\text{dip}}-\frac{1}{ic\mu_0}(a_1\vect{\mathrm{n}}_{1,1}+a_2\vect{\mathrm{n}}_{1,2}+b_1\vect{\mathrm{m}}_{1,1}+b_2\vect{\mathrm{m}}_{1,2})e^{-i\omega t},\\
    \vect{E}=\vect{E}_{\text{quad}}-(a_1\vect{\mathrm{m}}_{1,1}+a_2\vect{\mathrm{m}}_{1,2}+b_1\vect{\mathrm{n}}_{1,1}+b_2\vect{\mathrm{n}}_{1,2})e^{-i\omega t},\\
    \end{gathered}
    $}
\end{equation}
where $a_1$, $a_2$, $b_1$ and $b_2$ are constants to be determined by the boundary conditions (\ref{eqn:boundary_conditions_extended}). These constants may not necessarily be real valued.

\subsubsection{H Field Radial Component}
By substituting the radial components from (\ref{eqn:m_n_eval}) and (\ref{eqn:mag_dipole}) into \eqref{eqn:series_expansion} we have,
\begin{equation*}
\begin{aligned}
    H_r=&\frac{a^3}{ r^3}R_1(a)\cos\chi\cos\theta\\
    &- \frac{1}{ic\mu_0}\left(2a_1\frac{h_1(\rho)}{\rho}\sin\theta e^{i\varphi}+9a_2\frac{h_2(\rho)}{\rho}\sin{2\theta} e^{i\varphi}\right)e^{-i\omega t},
\end{aligned}
\end{equation*}
which, when $r=a$ ($\rho=\alpha$), will be equal to the boundary condition on $H_r$ from (\ref{eqn:boundary_conditions_extended}).
We can use $e^{i\varphi}e^{-i\omega t}=e^{i(\varphi-\omega t)}=e^{i\lambda}$ to write the exponent in terms of $\lambda$ and that $e^{i\lambda}=\cos\lambda+i\sin\lambda$. The constants can then be determined by matching trigonometric functions. For $\sin\theta$,
\begin{equation*}
    \Re\left(-\frac{1}{ic\mu_0}a_1\frac{h_1(\alpha)}{\alpha}e^{i\lambda}\right)=R_1(a)\sin\chi\cos\lambda,
\end{equation*}
where the $\Re$ operator has come from \eqref{eqn:oddeven}.
Hence, $a_1$ must be purely imaginary to cancel out $i$ in the bottom of the fraction so that $\cos\lambda$ is obtained instead of $\sin\lambda$.
\begin{gather}
    -\frac{2}{ic\mu_0}a_1\frac{h_1(\alpha)}{\alpha}\cos\lambda=R_1(a)\sin\chi\cos\lambda, \nonumber\\
    \implies a_1=-\frac{R_1(a)ic\mu_0\sin\chi}{2}\frac{\alpha}{h_1(\alpha)} \label{eqn:a_1}.
\end{gather}

As there is no $\sin{2\theta}$ term in the boundary condition, we will have $a_2=0$.
The gives the radial component of the $H$ field as the real part of
\begin{equation*}
    H_r=R_1(a)\left(\frac{a^3}{r^3}\cos\chi\cos\theta+\frac{\alpha}{h_1(\alpha)}\frac{h_1(\rho)}{\rho}\sin\chi\sin\theta e^{i\lambda}\right).
\end{equation*}

\subsubsection{E Field Polar Component}
Now considering the $E_\theta$ component which can be found by combing the $\angunitvect{\theta}$ components of results (\ref{eqn:m_n_eval}) and (\ref{eqn:ele_quadrupole}) into (\ref{eqn:series_expansion}),
\begin{equation*}
    \begin{aligned}
    E_\theta=&- \frac{1}{2}\omega \mu_0 a R_1(a)\frac{a^4}{r^4}\cos\chi\sin2\theta - a_1 i h_1(\rho) e^{i\varphi} e^{-i\omega t}\\
    &-b_1\frac{1}{\rho}(\rho h_1'(\rho)+h_1(\rho))\cos\theta e^{i\varphi}e^{-i\omega t}\\
    &-b_2\frac{3}{\rho}(\rho h_2'(\rho)+h_2(\rho))\cos{2\theta} e^{i\varphi}e^{-i\omega t},
    \end{aligned}
\end{equation*}
and when $r=a$ this is equal to the $E_\theta$ component from (\ref{eqn:boundary_conditions_extended}).
Once again, matching functions of $\theta$ and using $e^{i\varphi}e^{-i\omega t}=e^{i\lambda}$, we find that for
the term excluding any function of $\theta$,
\begin{equation*}
\begin{gathered}
    \Re\left(-\frac{c\mu_0R_1(a)}{2}\frac{\alpha}{h_1(\alpha)}h_1(\alpha)\sin\chi e^{i\lambda}\right)\\
    =-\frac{1}{2}\omega\mu_0aR_1(a)\sin\chi\cos\lambda,
\end{gathered}
\end{equation*}
then it is convenient to change $\alpha$ to $(\omega/c)a$,
\begin{equation*}
    -\frac{\omega\mu_0aR_1(a)}{2}\sin\chi\cos\lambda=-\frac{1}{2}\omega\mu_0aR_1(a)\sin\chi\cos\lambda,
\end{equation*} 
the LHS and RHS match, which validates the result for $a_1$ found in (\ref{eqn:a_1}). As the boundary condition does not contain any $\cos\theta$ terms,
\begin{equation}
\label{eqn:b_1}
\begin{gathered}
    -b_1\frac{1}{\rho}(\rho h_1'(\rho)+h_1(\rho))\cos\theta e^{i\lambda}=0,\\
    \implies b_1=0.
\end{gathered}
\end{equation}
Lastly, the $\cos{2\theta}$ terms,
\begin{gather*}
    \Re\left(-b_2\frac{3}{\alpha}(\alpha h_2'(\alpha)+h_2(\alpha))\cos{2\theta} e^{i\lambda}\right)\\
    =-\frac{1}{2}\omega\mu_0aR_1(a)\cos{2\theta}\sin\chi\cos\lambda,
\end{gather*}
again we want $\cos\lambda$ instead of $\sin\lambda$ so $b_2$ must be purely real. Hence,

\begin{gather}
-b_2\frac{3}{\alpha}(\alpha h_2'(\alpha)+h_2(\alpha))\cos{2\theta}\cos\lambda \nonumber \\
=\frac{1}{2}\omega\mu_0aR_1(a)\cos{2\theta}\sin\chi\cos\lambda, \nonumber \\
\implies b_2=-\frac{1}{2}\omega\mu_0aR_1(a)\frac{1}{3}\frac{\alpha}{\alpha h_2'(\alpha)+h_2(\alpha)}\sin\chi \label{eqn:b_2}.
\end{gather}
All the results for the constants thus far lead to,
\begin{equation}
\label{eqn:gen_E_theta}
\resizebox{\linewidth}{!}{$
\begin{aligned}
    E_\theta=&\frac{1}{2}\omega\mu_0 a R_1(a) \left[-\frac{a^4}{r^4}\cos\chi\sin{2\theta}\right.\\
    &+ \left.\left(\frac{\alpha}{\alpha h_2'(\alpha)+h_2(\alpha)}\frac{\rho h_2'(\rho) + h_2(\rho)}{\rho}\cos{2\theta} - \frac{h_1(\rho)}{h_1(\alpha)}\right)\sin\chi e^{i\lambda}\right]. 
\end{aligned}$
}
\end{equation}

\Deutschspace has a typographical error in his Appendix for eqn. \eqref{eqn:gen_E_theta}. The $\frac{\alpha}{\alpha h_2'(\alpha)+h_2(\alpha)}\frac{\rho h_2'(\rho) + h_2(\rho)}{\rho}$ term is instead written as $\frac{\alpha h_2'(\alpha)+h_2(\alpha)}{\alpha}\frac{\rho}{\rho h_2'(\rho)+h_2(\rho)}$.

\subsubsection{E Field Azimuthal Component}

The final boundary condition is on the $E_\varphi$ component. We combine the results from above with azimuthal component of (\ref{eqn:series_expansion}), which yields
\begin{equation*}
\resizebox{\linewidth}{!}{$
    \begin{aligned}
    E_\varphi=&-\left(-a_1 h_1(\rho)\cos\theta+b_2\frac{3i}{\rho}(\rho h_2'(\rho)+h_2(\rho))\cos\theta\right)e^{i\lambda}\\
    =&\frac{1}{2}\omega\mu_0aR_1(a) \left(\frac{\alpha}{\alpha h_2'(\alpha)+h_2(\alpha)}\frac{\rho h_2'(\rho)+h_2(\rho)}{\rho} - \frac{h_1(\rho)}{h_1(\alpha)}\right)i\sin\chi\cos\theta e^{i\lambda},
    \end{aligned}
    $}
\end{equation*}
and clearly at $r=a$ ($\rho=\alpha$) this reduces to zero and matches the required boundary condition $E_\varphi(a,\theta,\varphi)=0$, implying that the values for the constants are consist with all the boundary conditions in this situation.

\subsubsection{Remaining Vector Field Components}
Noting that the expressions for the constants $a_1$ and $b_2$ can be rewritten using $\alpha=(\omega/c)a$,  the field components are given by the real part of the following,
\begin{equation*}
\resizebox{\linewidth}{!}{$
    \begin{aligned}
    H_\theta&=\frac{1}{2}\frac{a^3}{ r^3}R_1(a)\cos\chi\sin\theta - \frac{1}{i c\mu_0}\left(a_1\frac{\rho h_1'(\rho)+h_1(\rho)}{\rho}\cos\theta + b_2(3ih_2(\rho)\cos\theta)\right)e^{i\lambda}\\
    &=\frac{1}{2}R_1(a)\left[\frac{a^3}{r^3}\cos\chi\sin\theta + \left(\frac{\alpha}{h_1(\alpha)}\frac{\rho h_1'(\rho)+h_1(\rho)}{\rho}\right.\right.\\
    &\qquad\qquad\qquad\qquad\qquad\qquad \left.\left. + \frac{1}{c}\omega a\frac{\alpha}{\alpha h_2'(\alpha)+h_2(\alpha)} h_2(\rho)\right)\sin\chi\cos\theta e^{i\lambda}\right]\\
    &=\frac{1}{2}R_1(a)\left[\frac{a^3}{r^3}\cos\chi\sin\theta+\left(\frac{\alpha}{h_1(\alpha)}\frac{\rho h_1'(\rho)+h_1(\rho)}{\rho} +\frac{\alpha^2h_2(\rho)}{\alpha h_2'(\alpha)+h_2(\alpha)} \right)\sin\chi\cos\theta e^{i\lambda}\right],\\
    \\
    H_\varphi&=-\frac{1}{i c\mu_0}\left(a_1\frac{i}{\rho}(\rho h_1'(\rho)+h_1(\rho)) - 3 b_2 h_2(\rho)\cos{2\theta}\right)e^{i\lambda}\\
    &=\frac{1}{2}R_1(a)\left(i\frac{\alpha}{h_1(\alpha)}\frac{\rho h_1'(\rho)+h_1(\rho)}{\rho}\sin\chi-\frac{1}{i}\frac{\alpha^2}{\alpha h_2'(\alpha)+h_2(\alpha)}h_2(\rho)\sin\chi\cos{2\theta}\right) e^{i\lambda}\\
    &=\frac{1}{2}R_1(a)\left(\frac{\alpha}{h_1(\alpha)}\frac{\rho h_1'(\rho)+h_1(\rho)}{\rho}+\frac{\alpha^2}{\alpha h_2'(\alpha)+h_2(\alpha)}h_2(\rho)\cos{2\theta}\right)i\sin\chi e^{i\lambda},\\
    \\
    E_r &= -\frac{1}{4}\omega\mu_0 a\frac{a^4}{r^4}R_1(a)\cos\chi\left(3\cos2\theta+1\right)-9b_2\frac{h_2(\rho)}{\rho}\sin{2\theta} e^{i\lambda}\\
    &=\frac{1}{2}\omega\mu_0aR_1(a)\left(-\frac{1}{2}\frac{a^4}{r^4}\cos\chi(3\cos{2\theta}+1)+3\frac{\alpha}{\alpha h_2'(\alpha)+h_2(\alpha)}\frac{h_2(\rho)}{\rho}\sin\chi\sin{2\theta} e^{i\lambda}\right).
    \end{aligned}
    $}
\end{equation*}
    
\begin{figure}[ht]
    \centering
    \includegraphics[width=0.8\textwidth]{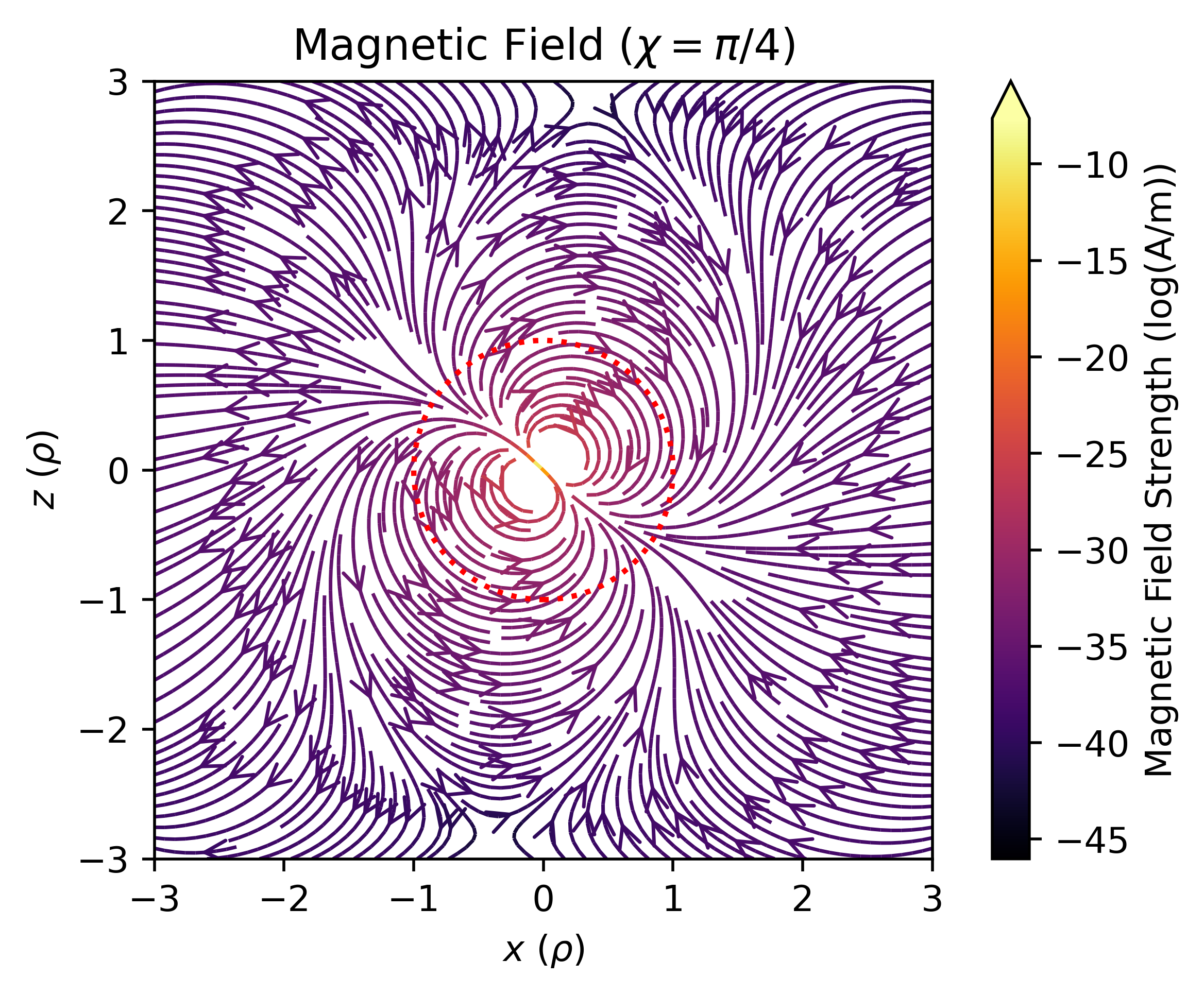}
    \caption{The magnetic field from \eqref{eqn:general_field_equations} using typical star parameters is plotted as streamlines showing the direction and the log magnitude of the magnetic field strength represented by the colour. Directly vertical is the star's rotation axis, and the magnetic field axis is set as $\chi=\pi/4$. Note the dashed red circle is the light cylinder, where $\rho=1$ or $r=c/\omega$.}
    \label{fig:Field_Large}
\end{figure}

\noindent The components of the general field equations are gathered in \ref{appendix:general_field_equations} for easier reference. A plot\footnote{For animations of the solution see \url{https://github.com/jsa113/General_Field_Equation_gifs}} of the magnetic field is shown in Fig. \ref{fig:Field_Large}. 

\section{The Radiation Fields of the Star}
\label{sec:radiation_fields}
\textcolor{black}{Far from a star, the electromagnetic fields take on their radiation form, allowing us to find the fields which `leak' out of the magnetosphere and propagate away from the electromagnetic source. To find the radiation form of the electromagnetic fields we consider the limit in which $r\gg c/\omega$ \citep{griffiths_introduction_2017}.} 
Hence, from \eqref{eqn:rho_def} if $r\gg c/\omega$ then $\rho\gg1$. Then the limit of the spherical Bessel functions when $\rho$ is large (\ref{eqn:bessel_fun_rho_big}) can be substituted into (\ref{eqn:general_field_equations}). 
\textcolor{black}{However, we find that $\alpha=(\omega/c)a\ll1$, which means $h_1(\alpha)$ can be approximated using (\ref{eqn:bessel_fun_rho_small}).} Using $c\simeq\SI{3e8}{\metre\per\second}$ and the typical star values given by \Deutsch,
\begin{equation*}
    \begin{gathered}
        a\simeq\SI{e9}{\metre},\quad \omega\simeq\SI{e-5}{rad \per s},
    \end{gathered}
\end{equation*}
we get $\alpha\simeq 3\times 10^{-5}\ll 1$. While a typical neutron star has 
\begin{equation*}
    \begin{gathered}
        a\simeq\SI{e4}{\metre},\quad \omega\simeq2\pi\,\SI{}{rad \per s},
    \end{gathered}
\end{equation*}
which gives $\alpha\simeq 2\times 10^{-4}\ll 1$. \textcolor{black}{We can also express this as a scaled equation,
\begin{equation*}
    \alpha = 2\times 10^{-4}\left(\frac{a}{\SI{e4}{m}}\right)\left(\frac{\omega}{2\pi\,\SI{}{rad \per s}}\right)
\end{equation*}
}
\textcolor{black}{The fastest recorded neutron star has $a\simeq\SI{1.6e4}{\metre}$ and $\omega\simeq\SI{4.5e3}{rad \per s}$ \citep{Hessels_2006}. We see from these values that $\alpha\simeq 0.2 < 1$. Though this limit is not as strong as slower neutron stars, it shows an appreciable upper limit that is still less than one. Suggesting that taking the approximation $\alpha\ll1$ in the following derivations is valid for slower stars but can still apply to faster stars, but with some error.}

\subsection{Radiation Form Derivation}
\textcolor{black}{To derive the radiation form we take the limit in which $r\gg c/\omega$. In taking this limit, we only consider terms that will be of the largest magnitude for each electromagnetic field component, neglecting terms containing high powers of $r$ in their denominator. Recall $\Re$ denotes the real part of an expression from \eqref{eqn:oddeven} and $\lambda=\varphi-\omega t$ from \eqref{eqn:lambda}.} 
Working firstly on the $H$ field's radial component, after neglecting smaller magnitude terms \textcolor{black}{and substituting the limit forms of $h_1(\rho)$ and $h_1(\alpha)$ in, $H_r$} becomes
\begin{equation*}
\begin{aligned}
    H_r&\simeq\Re\left({R_1(a)\frac{e^{i\rho}}{\rho^2}}\frac{\alpha^3}{i}\sin\chi\sin\theta e^{i\lambda}\right)\\
    &=\Re\left(R_1(a)\frac{\omega}{c}\frac{a^3}{r^2}\frac{1}{i}\sin\chi\sin\theta e^{i(\lambda+\rho)}\right)\\
    &=\Re\left(R_1(a)\frac{\omega}{c}\frac{a^3}{r^2}\frac{1}{i}\sin\chi\sin\theta\left[\cos(\lambda+\rho)+i\sin(\lambda+\rho)\right]\right)\\
    &=\frac{\omega}{c}a^3R_1(a)\frac{1}{r^2}\sin\chi\sin\theta\sin\left[\omega\left(\frac{r}{c}-t\right)+\varphi\right],
\end{aligned}   
\end{equation*}
\textcolor{black}{where $\lambda+\rho=\varphi-\omega t + \omega r/c=\omega(r/c-t)+\varphi$.
Repeating the same limit and substituting the approximations in for the polar component yields,}
\begin{equation*}
\resizebox{\linewidth}{!}{
    $\begin{aligned}
    H_\theta&\simeq\Re\left(\frac{1}{2}R_1(a)\left[\left(\frac{\alpha^2}{\alpha\frac{9i}{\alpha^4}-\frac{3i}{\alpha^3}}\right)\frac{ie^{i\rho}}{\rho}+\frac{\alpha^3}{-i}\left(\frac{-ie^{i\rho}}{\rho}-\frac{e^{i\rho}}{\rho^2}\right)\right]\sin\chi\cos\theta e^{i\lambda}\right)\\
    &=\Re\left(\frac{1}{2}R_1(a)\left[\frac{\alpha^5}{6\rho}+\frac{\alpha^3}{\rho}+\frac{\alpha^3}{i\rho^2}\right]\sin\chi\cos\theta e^{i(\lambda+\rho)}\right).
    \end{aligned}$
    }
\end{equation*}
The term with the largest magnitude in the line above is the $\alpha^3/\rho$ term as $\alpha\ll1$ and $\rho\gg1$. So,
\begin{equation*}
    \begin{aligned}
    H_\theta&\simeq\Re\left(\frac{1}{2}R_1(a)\frac{\alpha^3}{\rho}\sin\chi\cos\theta e^{i(\lambda+\rho)}\right)\\
    &=\Re\left(\frac{1}{2}R_1(a)\frac{\alpha^3}{\rho}\sin\chi\cos\theta \left[\cos(\lambda+\rho)+i\sin(\lambda+\rho)\right]\right)\\
    &=\frac{1}{2}\frac{\omega^2}{c^2}a^3R_1(a)\frac{1}{r}\sin\chi\cos\theta\cos\left[\omega\left(\frac{r}{c}-t\right)+\varphi\right].
    \end{aligned}
\end{equation*}
The azimuthal component has similar intermediate steps as above. Hence skipping intermediate working gives, 
\begin{equation*}
\resizebox{\linewidth}{!}{
    $\begin{aligned}
    H_\varphi&\simeq\Re\left(\frac{1}{2}R_1(a)\left[\left(\frac{\alpha^2}{\frac{9i}{\alpha^4}-\frac{3i}{\alpha^3}}\right)\frac{ie^{i\rho}}{\rho}\cos{2\theta}-\frac{\alpha^3}{i}\left(\frac{-ie^{i\rho}}{\rho}+\frac{-e^{i\rho}}{\rho^2}\right)\right]i\sin\chi e^{i\lambda}\right)\\
    &=\Re\left(\frac{1}{2}R_1(a)\left[\left(\frac{\alpha^6}{9-3\alpha}\right)\frac{1}{\rho}\cos{2\theta}+\frac{\alpha^3}{\rho}+\frac{\alpha^3}{i\rho^2}\right]i\sin\chi e^{i(\lambda+\rho)}\right).
    \end{aligned}$
    }
\end{equation*}
The largest term in the line above is the $\alpha^3/\rho$ term. Removing all smaller magnitude terms yields,
\begin{equation*}
    \begin{aligned}
    H_\varphi&\simeq\Re\left(\frac{1}{2}R_1(a)\frac{\alpha^3}{\rho}i\sin\chi e^{i(\lambda+\rho)}\right)\\
    &=\frac{1}{2}\frac{\omega^2}{c^2}a^3R_1(a)\frac{1}{r}\sin\chi\sin\left[\omega\left(\frac{r}{c}-t\right)+\varphi\right].
    \end{aligned}
\end{equation*}

Now consider the $E$ field. Using similar intermediate steps and the same principle as before where the largest order term is the only term considered \textcolor{black}{and the limit forms of $h_1(\rho)$ and $h_1(\alpha)$ are used}. The E~field's radial component will be,
\begin{equation*}
    \begin{aligned}
    E_r&\simeq\Re\left(\frac{1}{2}\omega\mu_0aR_1(a)3\left(\frac{\alpha}{\alpha\frac{9i}{\alpha^4}-\frac{3i}{\alpha^3}}\right)\frac{ie^{i\rho}}{\rho^2}\sin\chi\sin{2\theta}e^{i\lambda}\right)\\
    &=\Re\left(\frac{3}{2}\omega\mu_0aR_1(a)\frac{\alpha^4}{6}\frac{1}{\rho^2}\sin\chi\sin{2\theta}e^{i(\lambda+\rho)}\right)
    \end{aligned}
\end{equation*}
however, due to the $\alpha^4/\rho^2$ term, the radial component is equivalently equal to zero when compared to the other $E$ field components, this implies that $E_r\approx0$. The polar component is,
\begin{equation*}
\resizebox{\linewidth}{!}{
    $\begin{aligned}
    E_\theta&\simeq\Re\left(\frac{1}{2}\omega\mu_0aR_1(a)\left[\frac{\alpha^4}{6i}\left(\frac{-e^{i\rho}+\frac{ie^{i\rho}}{\rho}}{\rho}\right)\cos{2\theta}-\frac{\alpha^2}{i\rho}e^{i\rho}\right]\sin\chi e^{i\lambda}\right)\\
    &=\Re\left(\frac{1}{2}\omega\mu_0aR_1(a)\left[\frac{\alpha^4}{6i}\left(\frac{-1}{\rho}+\frac{i}{\rho^2}\right)\cos{2\theta}-\frac{\alpha^2}{i\rho}\right]\sin\chi e^{i(\lambda+\rho)}\right)\\
    &\simeq-\frac{1}{2}\frac{\omega^2\mu_0}{c}a^3R_1(a)\frac{1}{r}\sin\chi\sin\left[\omega\left(\frac{r}{c}-t\right)+\varphi\right].
    \end{aligned}$
    }
\end{equation*}
By using some of the previous intermediate steps from above, the azimuthal component is,
\begin{equation*}
\resizebox{\linewidth}{!}{
    $\begin{aligned}
    E_\varphi&\simeq\Re\left(\frac{1}{2}\omega\mu_0aR_1(a)\left[\frac{\alpha^4}{6i}\left(\frac{-1}{\rho}+\frac{i}{\rho^2}\right)-\frac{\alpha^2}{i\rho}\right]i\sin\chi\cos\theta e^{i(\lambda+\rho)}\right)\\
    &\simeq-\frac{1}{2}\frac{\omega^2\mu_0}{c}a^3R_1(a)\frac{1}{r}\sin\chi\cos\theta\cos\left[\omega\left(\frac{r}{c}-t\right)+\varphi\right].
    \end{aligned}$
    }
\end{equation*}
The results of the radiation field equations are gathered in \ref{appendix:radiation_field} as equations \eqref{eqn:mag_field_rad_form} and \eqref{eqn:ele_field_rad_form}. The results obtained here match the results obtained by \Deutsch.

These results are interesting as we can see the \emph{retarded time} affect, $(\frac{r}{c}-t)$, on the magnetic field in the radiation zone. That is we would not expect an instantaneous change to the electromagnetic fields when it is far from the surface from the star as electromagnetic ``news'' travels at the speed of light (\Griffiths). Instead we have a delayed affect which depends on the distance from the star. 

\subsection{Power Equation - D55 Equation (15)}
The power $P$ passing through a surface is the integral of the Poynting vector ($\vect{S}=\vect{E}\times \vect{H}$) over said surface. From \Griffiths,  Eqn.~(11.1),
\begin{equation}
    P(r,t)=\oint{\vect{S}\cdot d\vect{a}}.
\end{equation}
This is the power passing through the surface, so a negative must be introduced to give the rate at which energy is radiated away from the system,
\begin{equation}
\label{eqn:power_integral}
    -\frac{dW}{dt}=\oint{\vect{S}\cdot d\vect{a}}.
\end{equation}

For the $E$ and $H$ fields \Griffithsspace advocates taking the fields when $r$ is very large, when they are in \textit{radiation form}. Therefore, by taking the field equations (\ref{eqn:mag_field_rad_form}) and (\ref{eqn:ele_field_rad_form}) (which apply when $r\gg c/\omega$) leads to the Poynting vector,
\begin{equation*}
    \vect{S}=\begin{pmatrix}
        0\\
        E_{\theta}\\
        E_{\varphi}
        \end{pmatrix}
        \times
        \begin{pmatrix}
        H_r\\
        H_{\theta}\\
        H_{\varphi}
        \end{pmatrix}
        =\begin{pmatrix}
        E_{\theta}H_{\varphi}-E_{\varphi}H_{\theta}\\
        E_{\varphi}H_r\\
        E_{\theta}H_r
    \end{pmatrix}
\end{equation*}
and by also taking the surface to be a sphere of radius $r$, this implies that $S_r\parallel\vect{a}$, $S_\theta\perp\vect{a}$ and $S_\varphi\perp\vect{a}$, as $\vect{a}$ is the normal of the surface (see Fig. \ref{fig:polar_vector_field}). 
So (\ref{eqn:power_integral}) becomes,
\begin{equation*}
    -\frac{dW}{dt}=\int_0^{2\pi}\int_0^\pi(E_{\theta}H_{\varphi}-E_{\varphi}H_{\theta})r^2\sin(\theta)\,d\theta\,d\varphi.
\end{equation*}
Upon applying expressions for the $H$ and $E$ fields from (\ref{eqn:mag_field_rad_form}) and (\ref{eqn:ele_field_rad_form}) we get,

\begin{equation}
\label{eqn:(15)_int_form}
\resizebox{\linewidth}{!}{$
\begin{aligned}
    -\frac{dW}{dt}=\int_0^{2\pi}\int_0^\pi &\frac{1}{4}\frac{\omega^4\mu_0}{c^3}a^6 R_1^2 (a)\sin^2(\chi)\left\{\sin^2\left[\omega\left(\frac{r}{c} - t\right) + \varphi\right]\right.\\
    & \left. + \cos^2(\theta)\cos^2\left[\omega\left(\frac{r}{c} - t\right) + \varphi\right]\right\}\sin(\theta)\,d\theta\,d\varphi.
\end{aligned}
$}
\end{equation}

Now we evaluate the integrals over the functions that depend on $\theta$ and $\varphi$. The double integral can firstly be expressed as,
\begin{gather}
    \begin{aligned}\label{eqn:trig_power_integral}
        \int_0^{2\pi}\int_0^\pi &\sin(\theta)\sin^2\left[\omega\left(\frac{r}{c}-t\right)+\varphi\right]\\ &+\sin(\theta)\cos^2(\theta)\cos^2\left[\omega\left(\frac{r}{c}-t\right)+\varphi\right]\,d\theta\,d\varphi,
    \end{aligned}\\
    \shortintertext{and using the integration identities,}
    \int_0^{2\pi}\sin(\theta)\,d\theta =2, \nonumber\\
    \int_0^{2\pi}\sin(\theta)\cos^2(\theta)\,d\theta=\frac{2}{3}, \nonumber\\
    \begin{gathered}
        \int_0^\pi\sin^2\left[\omega\left(\frac{r}{c}-t\right)+\varphi\right]\,d\varphi\\
        =\int_0^\pi\cos^2\left[\omega\left(\frac{r}{c}-t\right)+\varphi\right]\,d\varphi=\pi,
    \end{gathered} \nonumber
\end{gather}
which leads to \emph{the integral} \eqref{eqn:trig_power_integral} yielding $8\pi/3$. On combining this result with (\ref{eqn:(15)_int_form}) we get
\begin{equation}
\label{eqn:energy_radiated}
    -\frac{dW}{dt}=\frac{2\pi\omega^4\mu_0}{3c^3}a^6R_1^2(a)\sin^2(\chi).
\end{equation}
This is the same result obtained by \Deutschspace in his equation (15). The RHS is always positive. Hence, $dW/dt$ must always be negative. That is, energy is always being radiated away from the star. The energy that is radiated comes from the rotational kinetic energy causing the star's spin down.



\subsubsection{Spin Down Rate - D55 Equation (16)}
To derive the spin down rate of a star in the VRD solution, we require two standard relationships  (see for example \cite{moebs_university_2016}, Eqns. (10.31) and (11.8)),
\begin{equation*}
    \vect{\tau}=\frac{d\vect{L}}{dt}, \qquad \frac{dW}{dt}=P=\vect{\tau}\cdot\vect{\omega}, \quad \implies \quad \frac{dW}{dt}=\frac{d\vect{L}}{dt}\cdot\vect{\omega},
\end{equation*}
where $\vect{L}$ is the angular momentum and $\vect{\tau}$ is the torque.
As the angular velocity and the angular momentum are in the same direction, we can rearrange the equation above to become,
\begin{equation*}
    \frac{d\vect{L}}{dt}=\frac{\vect{\omega}}{\norm{\vect{\omega}}^2}\frac{dW}{dt},
\end{equation*}
which is the result obtained by \Deutsch. 

However, another typo is present in \Deutsch. It is not obviously shown that his $\omega$ is actually a vector and that the $\omega$ in the bottom of the fraction is the norm of $\omega$. Explaining why he has $\omega/\omega^2$ rather than just $1/\omega$. 
Lastly, \Deutschspace displays it with a negative so it can be compared easier with (\ref{eqn:energy_radiated}),
\begin{equation}
\label{eqn:ang_mom_rad}
    -\frac{d\vect{L}}{dt}=\frac{\vect{\omega}}{\norm{\vect{\omega}}^2}\left(-\frac{dW}{dt}\right).
\end{equation}
It can be seen from this equation that the angular momentum also decreases with time. The star will be brought to rest by its own radiation as \Deutschspace states.

\section{Conclusion}
By considering an idealised star and assuming a dipolar form for the internal magnetic field, the general field equations of the electromagnetic fields external of the star were found analytically. These general field equations were then able to be reduced to their radiation limit ($r\gg c/\omega$). From the radiation form, the relationship for the power radiated by the rotating magnetic field was found. 

It was found that the power equation matches the equation found in \Deutschspace and the derivation for which is shown in this report for scrutiny. However, a typo appears in \Deutsch's general field equations. In particular, the $E_\theta$ \eqref{eqn:gen_E_theta} component appears to evaluate the wrong fraction at $\alpha$. This error in \Deutschspace causes trouble when reducing the general field equations to their radiation form. When using the $E_\theta$ component found in this report the reduction to radiation form matches \Deutsch.

A plot of the magnetic field using the general field equations was also produced, showing the summation of the magnetic dipole and the wave solution. This generalises a similar plot given in \Deutschspace which was zoomed in around the $r\ll c/\omega$ region.

\textcolor{black}{A 1969 paper by Goldreich and Julian argues that a pulsar cannot be surrounded by a vacuum but rather a magnetosphere containing plasma \citep{1969ApJ...157..869G}. 
A related approach} can be taken when estimating the power equation. The Force-Free (FF) magnetosphere, which is the solution to Maxwell’s Equations for a rotating star with a dipole field. It ignores the plasma pressure and requires that the electric field parallel to the magnetic field is zero everywhere \citep{contopoulos_axisymmetric_1999}. This solution requires approximations just like the VRD solution. The most realistic way is to use numerical simulations. The results from \cite{kalapotharakos_three-dimensional_2018} find their power equations ranging from the near FF solution to near the VRD solution given in equation \eqref{eqn:energy_radiated}. The VRD solution proves to be an interesting limiting case of numerical simulations.

\textcolor{black}{The VRD solution discussed in this review has been extended in a number of ways. For example,
general relativistic corrections to the Deustch solution can be added  \citep{10.1046/j.1365-8711.2001.04161.x,Petri2013}. Also, considering an  electromagnetic field that consists of more multipole terms than just a dipole adds extra correction terms that are potentially required to match reality \citep{10.1093/mnras/stv598, 2020arXiv200702539B}.
}




\bibliography{VRD_references}

\begin{thebibliography}{}
\expandafter\ifx\csname natexlab\endcsname\relax\def\natexlab#1{#1}\fi

\bibitem[{{Bonazzola} {et~al.}(2020){Bonazzola}, {Mottez}, \&
  {Heyvaerts}}]{2020arXiv200702539B}
{Bonazzola}, S., {Mottez}, F., \& {Heyvaerts}, J. 2020, arXiv e-prints,
  arXiv:2007.02539

\bibitem[{Contopoulos {et~al.}(1999)Contopoulos, Kazanas, \&
  Fendt}]{contopoulos_axisymmetric_1999}
Contopoulos, I., Kazanas, D., \& Fendt, C. 1999, The Astrophysical Journal,
  511, 351, arXiv: astro-ph/9903049

\bibitem[{Deutsch(1955)}]{deutsch_electromagnetic_1955}
Deutsch, A.~J. 1955, Annales d'Astrophysique, 18

\bibitem[{{Goldreich} \& {Julian}(1969)}]{1969ApJ...157..869G}
{Goldreich}, P., \& {Julian}, W.~H. 1969, \apj, 157, 869

\bibitem[{Griffiths(2017)}]{griffiths_introduction_2017}
Griffiths, D.~J. 2017, Introduction to {Electrodynamics}, 4th edn. (Cambridge
  University Press), doi:\url{10.1017/9781108333511}

\bibitem[{Hessels {et~al.}(2006)Hessels, Ransom, Stairs, Freire, Kaspi, \&
  Camilo}]{Hessels_2006}
Hessels, J. W.~T., Ransom, S.~M., Stairs, I.~H., {et~al.} 2006, Science, 311,
  1901

\bibitem[{Kalapotharakos {et~al.}(2018)Kalapotharakos, Brambilla, Timokhin,
  Harding, \& Kazanas}]{kalapotharakos_three-dimensional_2018}
Kalapotharakos, C., Brambilla, G., Timokhin, A., Harding, A.~K., \& Kazanas, D.
  2018, The Astrophysical Journal, 857, 44

\bibitem[{Melatos(1997)}]{melatos_spin-down_1997}
Melatos, A. 1997, Monthly Notices of the Royal Astronomical Society, 288, 1049

\bibitem[{Michel \& Li(1999)}]{michel_electrodynamics_1999}
Michel, F., \& Li, H. 1999, Physics Reports, 318, 227

\bibitem[{Michel(1991)}]{michel_theory_1991}
Michel, F.~C. 1991, Theory of neutron star magnetospheres, Theoretical
  astrophysics (Chicago: University of Chicago Press)

\bibitem[{Moebs {et~al.}(2016)Moebs, Ling, \& Sanny}]{moebs_university_2016}
Moebs, W., Ling, S.~J., \& Sanny, J. 2016, University {Physics} {Volume} 1
  (Houston, Texas: OpenStax)

\bibitem[{{P{\'e}tri}(2013)}]{Petri2013}
{P{\'e}tri}, J. 2013, \mnras, 433, 986

\bibitem[{Pétri(2015)}]{10.1093/mnras/stv598}
Pétri, J. 2015, Monthly Notices of the Royal Astronomical Society, 450, 714

\bibitem[{Rezzolla {et~al.}(2001)Rezzolla, Ahmedov, \&
  Miller}]{10.1046/j.1365-8711.2001.04161.x}
Rezzolla, L., Ahmedov, B.~J., \& Miller, J.~C. 2001, Monthly Notices of the
  Royal Astronomical Society, 322, 723

\bibitem[{Roberts(2007)}]{roberts_alfvens_2007}
Roberts, P.~H. 2007, in Encyclopedia of {Geomagnetism} and {Paleomagnetism},
  ed. D.~Gubbins \& E.~Herrero-Bervera (Dordrecht: Springer Netherlands), 7--11

\bibitem[{Stratton(1941)}]{stratton_electromagnetic_1941}
Stratton, J.~A. 1941, Electromagnetic {Theory}, International series in physics
  (McGraw-Hill book Company, Incorporated)

\bibitem[{Travelle(2011)}]{travelle_pulsars_2011}
Travelle, P.~A., ed. 2011, Pulsars: discoveries, functions, and formation
  (Hauppauge, N.Y: Nova Science Publishers)

\bibitem[{Van~Brummelen(2012)}]{van_brummelen_heavenly_2012}
Van~Brummelen, G. 2012, Heavenly {Mathematics} (Princeton University Press)

\end{thebibliography}

\appendix

\newpage
\begin{strip}
\section{General Field Equations}
\label{appendix:general_field_equations}
The results of the general field equations found in Sec. \ref{sec:general_field_equations} are gathered here for reference. The fields external a star for the VRD solution is the \emph{real} part of the following,
\begin{equation}
\label{eqn:general_field_equations}
\begin{aligned}
    H_r&=R_1(a)\left(\frac{a^3}{r^3}\cos\chi\cos\theta+\frac{\alpha}{h_1(\alpha)}\frac{h_1(\rho)}{\rho}\sin\chi\sin\theta e^{i\lambda}\right),\\
    H_\theta&=\frac{1}{2}R_1(a)\left[\frac{a^3}{r^3}\cos\chi\sin\theta+\left(\frac{\alpha}{h_1(\alpha)}\frac{\rho h_1'(\rho)+h_1(\rho)}{\rho}  + \frac{\alpha^2}{\alpha h_2'(\alpha)+h_2(\alpha)}h_2(\rho)\right)\sin\chi\cos\theta e^{i\lambda}\right],\\
    H_\varphi&=\frac{1}{2}R_1(a)\left(\frac{\alpha}{h_1(\alpha)}\frac{\rho h_1'(\rho)+h_1(\rho)}{\rho}+\frac{\alpha^2}{\alpha h_2'(\alpha)+h_2(\alpha)}h_2(\rho)\cos{2\theta}\right)i\sin\chi e^{i\lambda},\\
    E_r&=\frac{1}{2}\omega\mu_0aR_1(a)\left(-\frac{1}{2}\frac{a^4}{r^4}\cos\chi(3\cos{2\theta}+1)+3\frac{\alpha}{\alpha h_2'(\alpha)+h_2(\alpha)}\frac{h_2(\rho)}{\rho}\sin\chi\sin{2\theta} e^{i\lambda}\right),\\
    E_\theta&=\frac{1}{2}\omega\mu_0aR_1(a)\left[-\frac{a^4}{r^4}\cos\chi\sin{2\theta} + \left(\frac{\alpha}{\alpha h_2'(\alpha)+h_2(\alpha)}\frac{\rho h_2'(\rho) + h_2(\rho)}{\rho}\cos{2\theta} - \frac{h_1(\rho)}{h_1(\alpha)}\right)\sin\chi e^{i\lambda}\right],\\
    E_\varphi&=\frac{1}{2}\omega\mu_0aR_1(a)\left(\frac{\alpha}{\alpha h_2'(\alpha)+h_2(\alpha)}\frac{\rho h_2'(\rho)+h_2(\rho)}{\rho}-\frac{h_1(\rho)}{h_1(\alpha)}\right)i\sin\chi\cos\theta e^{i\lambda}. 
\end{aligned}
\end{equation}
\noindent\rule{\textwidth}{0.4pt}
\end{strip}
\section{Radiation Field Equations}
\label{appendix:radiation_field}
The results of the radiation field equations found in Sec. \ref{sec:radiation_fields} are gathered here for reference.
The magnetic field when $r\gg c/\omega$ is,
\begin{equation}
\label{eqn:mag_field_rad_form}
    \begin{aligned}
    H_r&=\frac{\omega}{c}a^3R_1(a)\frac{1}{r^2}\sin\chi\sin\theta\sin\left[\omega\left(\frac{r}{c}-t\right)+\varphi\right],\\
    H_\theta&=\frac{1}{2}\frac{\omega^2}{c^2}a^3R_1(a)\frac{1}{r}\sin\chi\cos\theta\cos\left[\omega\left(\frac{r}{c}-t\right)+\varphi\right],\\
    H_\varphi&=\frac{1}{2}\frac{\omega^2}{c^2}a^3R_1(a)\frac{1}{r}\sin\chi\sin\left[\omega\left(\frac{r}{c}-t\right)+\varphi\right].
    \end{aligned}
\end{equation}
The electric field when $r\gg c/\omega$ is,
\begin{equation}
\label{eqn:ele_field_rad_form}
    \begin{aligned}
    E_r&=0,\\
    E_\theta&=-\frac{1}{2}\frac{\omega^2\mu_0}{c}a^3R_1(a)\frac{1}{r}\sin\chi\sin\left[\omega\left(\frac{r}{c}-t\right)+\varphi\right],\\
    E_\varphi&=-\frac{1}{2}\frac{\omega^2\mu_0}{c}a^3R_1(a)\frac{1}{r}\sin\chi\cos\theta\cos\left[\omega\left(\frac{r}{c}-t\right)+\varphi\right].
    \end{aligned}
\end{equation}

\section{Analysis of Spherical Bessel Functions}
\label{appendix:bessel_func}

The general solution for the fields includes spherical Bessel functions of the third kind (first kind Hankel functions). This appendix highlights what these functions look like and what they are approximately equal to when considered at the limits $\rho\ll1$ and $\rho\gg1$. From Ref.~\cite{stratton_electromagnetic_1941} page 405,
\begin{equation}
\label{eqn:bessel_fun}
\begin{gathered}
    h_1^{(1)}(\rho) = -e^{i\rho}\left(\frac{\rho+i}{\rho^2}\right),\\
    h_2^{(1)}(\rho) = i e^{i\rho}\left(\frac{\rho^2+3i\rho-3}{\rho^3}\right).
\end{gathered}
\end{equation}
We will omit the superscript denoting the functions as the third kind for convenience. Differentiating these functions with respect to $\rho$ yields,
\begin{equation}
\label{eqn:bessel_fun_der}
\begin{gathered}
    h_1'=-e^{i\rho}\left(\frac{i\rho^2-2\rho-2i}{\rho^3}\right),\\
    h_2'=-e^{i\rho}\left(\frac{\rho^3+4i\rho^2-9\rho-9i}{\rho^4}\right).
\end{gathered}
\end{equation}
In the limit $\rho\ll1$ eqns. \eqref{eqn:bessel_fun} and \eqref{eqn:bessel_fun_der} become,
\begin{equation}
\label{eqn:bessel_fun_rho_small}
    \begin{aligned}
        h_1\approx&\frac{-i}{\rho^2},\qquad & h_2\approx&\frac{-3i}{\rho^3},\\
        h_1'\approx&\frac{2i}{\rho^3}, & h_2'\approx&\frac{9i}{\rho^4}.
    \end{aligned}
\end{equation}
In the limit $\rho\gg1$ eqns. \eqref{eqn:bessel_fun} and \eqref{eqn:bessel_fun_der} become,
\begin{equation}
\label{eqn:bessel_fun_rho_big}
    \begin{aligned}
        h_1\approx&\frac{-e^{i\rho}}{\rho},\qquad & h_2\approx&\frac{ie^{i\rho}}{\rho},\\
        h_1'\approx&\frac{-ie^{i\rho}}{\rho}, & h_2'\approx&\frac{-e^{i\rho}}{\rho}.
    \end{aligned}
\end{equation}

The form these Bessel functions take at these limits are relatively simple and are required in the derivation of a eqns. (13), (14), (18) and (19) from \Deutschspace (equations (\ref{eqn:mag_field_rad_form}) and (\ref{eqn:ele_field_rad_form}) in this report).

\section{Associated Legendre Functions}
\label{appendix:leg_fun}
In Appendix IV of \cite{stratton_electromagnetic_1941}, the associated Legendre functions are given as,
\begin{equation}
\label{eqn:ass_leg_fun}
    \begin{aligned}
    P_0(\cos\theta)&=1,\\
    P_1(\cos\theta)&=\cos\theta,\\
    P_1^1(\cos\theta)&=\sin\theta,\\
    P_2(\cos\theta)&=\frac{1}{2}(3\cos^2\theta-1)=\frac{1}{4}(3\cos{2\theta}+1),\\
    P_2^1(\cos\theta)&=3\cos\theta\sin\theta=\frac{3}{2}\sin{2\theta},
    \end{aligned}
\end{equation}
where $\theta\in[0,\pi]$, the subscript denotes the degree of the polynomial, and the superscript denotes the order of the derivative.

\section{Vector Calculus in Spherical Coordinates}
\label{appendix:spherical_calculus}

Let $f(r,\theta,\varphi)$ be a scalar function and $\vect{A}(r,\theta,\varphi)=(A_r,A_\theta,A_\varphi)$ be a vector-valued function in spherical coordinates $(r,\theta,\varphi)$. From appendix A of \Griffiths, the gradient of a scalar function is,
\begin{equation}
\label{eqn:gradient_spherical}
    \nabla f=\frac{\partial f}{\partial r}\unitvect{r}+\frac{1}{r}\frac{\partial f}{\partial \theta}\angunitvect{\theta}+\frac{1}{r\sin\theta}\frac{\partial f}{\partial \varphi}\angunitvect{\varphi}.
\end{equation}
The divergence of a vector field is,
\begin{equation}
\label{eqn:divergence_spherical}
\resizebox{\linewidth}{!}{$
    \nabla\cdot\vect{A}=\frac{1}{r^2}\frac{\partial}{\partial r}\left(r^2A_r\right)+\frac{1}{r\sin\theta}\frac{\partial}{\partial\theta}\left(A_\theta\sin\theta\right)+\frac{1}{r\sin\theta}\frac{\partial A_\varphi}{\partial\varphi}.
    $}
\end{equation}
The curl of a vector field is,
\begin{equation}
\label{eqn:curl_spherical}
    \begin{aligned}
    \nabla\times\vect{A}=&\frac{1}{r \sin \theta} \left(\frac{\partial}{\partial \theta}\left(A_{\varphi} \sin \theta\right)-\frac{\partial A_{\theta}}{\partial \varphi}\right)\unitvect{r}\\
    &+\frac{1}{r}\left(\frac{1}{\sin \theta} \frac{\partial A_{r}}{\partial \varphi}-\frac{\partial}{\partial r}\left(r A_{\varphi}\right)\right)\angunitvect{\theta}\\
    &+\frac{1}{r}\left(\frac{\partial}{\partial r}\left(r A_{\theta}\right)-\frac{\partial A_{r}}{\partial \theta}\right)\angunitvect{\varphi}.
    \end{aligned}
\end{equation}

\section{Solutions of Spherical Triangles}
\label{Appendix:SphericalTrangle}
From \cite{van_brummelen_heavenly_2012} on page 98 the spherical Law of Cosines is given as,
\begin{equation}
\label{eqn:sph_tri_2_sides}
    \cos c = \cos a \cos b + \sin a \sin b \cos C,
\end{equation}
and on page 63 the spherical Law of Sines is given as,
\begin{equation}
\label{eqn:sphi_tri_sin}
    \frac{\sin a}{\sin A} = \frac{\sin b}{\sin B}.
\end{equation}
Upon rearranging the spherical Law of Cosines for $\cos C$ we get,
\begin{equation}
\label{eqn:sph_tri_3_sides}
    \cos C=\frac{\cos c - \cos a\cos b}{\sin a \sin b}.
\end{equation}

\def\Rotation{160}
\def\Tilt{15}
\def\RadiusSphere{2}
\def\VectorLen{1.5}

\def\VectorFieldPolar{45}
\def\VectorFieldAzimuth{45}

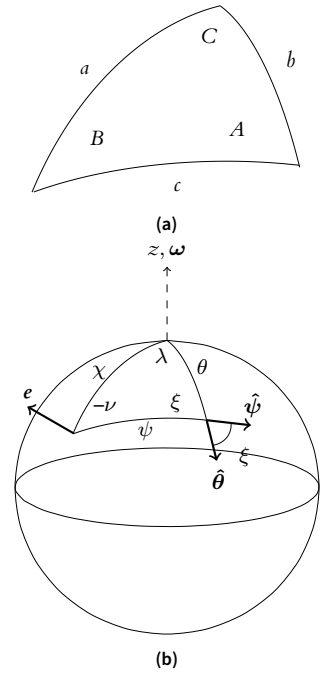
\begin{figure}[ht]
    \begin{subfigure}{0.5\textwidth}
        \centering
        \begin{tikzpicture}
            \begin{scope}[viewport={\Rotation}{\Tilt}, very thin]
                \draw[domain=0:58.5, variable=\polar, smooth] plot (\ToXYZr{2}{\polar}{\Rotation-136.5}) node[] at (\ToXYZr{2.2}{35}{\Rotation-135}) {$a$};
                
                \draw[domain=0:50, variable=\polar, smooth] plot (\ToXYZr{2}{\polar}{\Rotation-70}) node[] at (\ToXYZr{2.2}{30}{\Rotation-65}) {$b$};
                
                \node at (\ToXYZr{2}{41}{\Rotation-85}) {$A$};
                \node at (\ToXYZr{2}{45}{\Rotation-125}) {$B$};
                \node at (\ToXYZr{2}{15}{\Rotation-98}) {$C$};
            \end{scope}
            
            \tdplotsetmaincoords{0}{0}
            \tdplotsetrotatedcoords{-45+\Tilt}{-45}{0}
            \begin{scope}[tdplot_rotated_coords, very thin]
                \draw[domain=0:53, variable=\polar, smooth] plot (\ToXYZr{4}{\polar}{40}) node[] at (\ToXYZr{4}{30}{30}) {$c$};
            \end{scope}
            
        \end{tikzpicture}               
        \caption{}
        \label{fig:Spherical_Triangle_Appendix}
    \end{subfigure}
    \begin{subfigure}{0.5\textwidth}
        \centering
        \begin{tikzpicture}
            \begin{scope}[viewport={\Rotation}{\Tilt}, very thin]
            
                \draw[dashed,->] (0,0,1) -- (0,0,\VectorLen) node[anchor=south]{$z, \vect{\omega}$};
                
                \draw[domain=0:360, variable=\azimuth, smooth] plot (\ToXYZ{90}{\azimuth});
                \draw[domain=0:360, variable=\elevation, smooth] plot (\ToXYZ{\elevation}{\Rotation});
                
                \draw[domain=0:58.5, variable=\polar, smooth] plot (\ToXYZr{1}{\polar}{\Rotation-136.5}) node[] at (\ToXYZr{1.1}{35}{\Rotation-135}) {$\chi$};
                
                \draw[domain=0:50, variable=\polar, smooth] plot (\ToXYZr{1}{\polar}{\Rotation-70}) node[] at (\ToXYZr{1.1}{30}{\Rotation-65}) {$\theta$};
                
                \node at (\ToXYZr{1}{45}{\Rotation-125}) {$\tiny -\nu$};
                \node at (\ToXYZr{1}{41}{\Rotation-85}) {$\tiny\xi$};
                \node at (\ToXYZr{1}{15}{\Rotation-98}) {$\small\lambda$};
                
                \draw[domain=0:0.3, variable=\t, thick, ->] plot (\ThetaTangentGradient{1}{49.5}{\Rotation-70}{\t}) node[below]{$\angunitvect{\theta}$};

            \end{scope}
            \tdplotsetmaincoords{0}{0}
            \tdplotsetrotatedcoords{-45+\Tilt}{-45}{0}
            
            \def\EVectorFieldPolar{40}
            \def\EVectorFieldAzimuth{225}
            
            \begin{scope}[tdplot_rotated_coords, very thin]
                \draw[domain=2:3, variable=\radius, thick, ->] plot (\ToXYZr{\radius}{0}{0}) node[anchor=south]{$\vect{e}$};
                
                \draw[domain=0:53, variable=\polar, smooth] plot (\ToXYZr{2}{\polar}{40}) node[] at (\ToXYZr{2}{30}{30}) {$\psi$};
                
                \draw[domain=0:0.3, variable=\t, thick, ->] plot (\ThetaTangentGradient{2}{53}{40}{\t}) node[above]{$\angunitvect{\psi}$};
            \end{scope}

            \tdplotsetrotatedcoords{44.5+\Tilt}{30.5}{0}
            \begin{scope}[tdplot_rotated_coords, very thin]
                \draw[domain=222:300, variable=\azimuth, smooth] plot (\ToXYZr{2}{10}{\azimuth}) node[] at (\ToXYZr{2}{20}{261}) {$\xi$};
            \end{scope}
            
        \end{tikzpicture}     
        \caption{}
        \label{fig:Spherical Triangle}
    \end{subfigure}
    \caption{Spherical triangles with general notation and the labels according to parameters defined in the report. (a) Spherical Triangle with side lengths and angles labelled according to the notation in this appendix. Note that sides and angles labelled with the same letter are opposite each other. (b) Description according to coordinate parameters and inclination of $\vect{e}$ axis. Note that $\nu$ is an angle which is $2\pi$ periodic so $-\nu \equiv 2\pi-\nu$ (note angles increase though anticlockwise).}
\end{figure}

\end{document}